\documentclass[letterpaper,twocolumn,10pt]{article}
\usepackage{usenix}

\usepackage{graphicx} 
\usepackage[utf8]{inputenc}
\usepackage{tikz}
\usepackage{xspace}
\usepackage{xcolor}
\usepackage{xstring}
\usepackage{color, colortbl}
\usepackage{multicol,multirow}

\usepackage{enumitem}
\usepackage{subcaption}
\usepackage{fontawesome}
\usepackage{amsmath}
\usepackage{bm}
\usepackage{nicematrix}
\usepackage[pangram]{blindtext}
\usepackage{balance}
\usepackage{totpages}

\usepackage{listings}
\usepackage{multirow}
\usepackage{booktabs}
\usepackage{makecell}
\usepackage{enumitem}
\usepackage{array} 
\usepackage{seqsplit}
\usepackage{float}
\usepackage{diagbox}
\usepackage{threeparttable}
\usepackage{cleveref}
\usepackage{microtype}

\usepackage[framemethod=tikz]{mdframed}
\usepackage{tabularx}
\usepackage{tcolorbox}

\usepackage{pifont}
\usepackage{xcolor}

\usepackage{authblk}

\newcommand{\ie}{\textit{i}.\textit{e}.}
\newcommand{\eg}{\textit{e}.\textit{g}.}

\def\Snospace~{\S{}}

\newif\ifdraft\drafttrue
\newif\ifnotes\notestrue
\ifdraft\else\notesfalse\fi

\input{glyphtounicode}
\newcommand{\squishlist}{
\begin{itemize}[noitemsep,nolistsep]
  \setlength{\itemsep}{-0pt}
}
\newcommand{\squishend}{
  \end{itemize}
}

\newcommand{\PP}[1]{
\noindent{\bf \IfEndWith{#1}{.}{#1}{#1.}}
}

\newcommand{\PPNS}[1]{
\noindent{\bf \IfEndWith{#1}{.}{#1}{#1.}}
}

\newcommand{\boxbeg}{
\noindent\begin{tabular}{|l|}\hline
\begin{minipage}{3.2in}
\noindent
}

\newcommand{\boxend}{
\end{minipage}\\ \hline
\end{tabular}
}

\newcommand{\aref}[1]{\hyperref[#1]{Appendix~\ref*{#1}}}

\newcommand*\justify{%
  \fontdimen2\font=0.4em
  \fontdimen3\font=0.2em
  \fontdimen4\font=0.1em
  \fontdimen7\font=0.1em
  \hyphenchar\font=`\-
}

\renewcommand{\texttt}[1]{%
  \begingroup
  \ttfamily
  \begingroup\lccode`~=`/\lowercase{\endgroup\def~}{/\discretionary{}{}{}}%
  \begingroup\lccode`~=`[\lowercase{\endgroup\def~}{[\discretionary{}{}{}}%
  \begingroup\lccode`~=`.\lowercase{\endgroup\def~}{.\discretionary{}{}{}}%
  \catcode`/=\active\catcode`[=\active\catcode`.=\active
  \justify\scantokens{#1\noexpand}%
  \endgroup
}

\newcolumntype{P}[1]{>{\centering\arraybackslash}p{#1}}

\newtcolorbox{boxI}{
    colback = lightgray!10, 
    colframe = black, 
    boxrule = 0.5pt, 
    toprule = 0.5pt, 
    arc = 2pt,
    left = 1pt,
    right = 1pt,
    bottom = 0pt,
    top = 0pt
}

\newcounter{observcntr}
\newcommand*{\observ}[1]{%
    \stepcounter{observcntr}%
    \begin{center}
        \begin{boxI}
        \textbf{Observation~\arabic{observcntr}: }{#1}. 
        \end{boxI}
    \end{center}
}
\begin{document}

\title{\Large \bf How Can We Effectively Use LLMs for Phishing Detection?: Evaluating the \\Effectiveness of Large Language Model-based Phishing Detection Models}


\author{Fujiao Ji, Doowon Kim}
\affil{\textit{University of Tennessee, Knoxville}}

\maketitle

\begin{abstract}
Large language models (LLMs) have emerged as a promising phishing detection mechanism, addressing the limitations of traditional deep learning-based detectors, including poor generalization to previously unseen websites and a lack of interpretability. However, LLMs' effectiveness for phishing detection remains unexplored. This study investigates how to effectively leverage LLMs for phishing detection (including target brand identification) by examining the impact of input modalities (screenshots, logos, HTML, and URLs), temperature settings, and prompt engineering strategies. Using a dataset of 19,131 real-world phishing websites and 243 benign sites, we evaluate seven LLMs -- two commercial models (GPT 4.1 and Gemini 2.0 flash) and five open-source models (Qwen, Llama, Janus, DeepSeek-VL2, and R1) -- alongside two deep learning (DL)-based baselines (PhishIntention and Phishpedia).

Our findings reveal that commercial LLMs generally outperform open-source models in phishing detection, while DL models demonstrate better performance on benign samples. For brand identification, screenshot inputs achieve optimal results, with commercial LLMs reaching 93-95\% accuracy and open-source models, particularly Qwen, achieving up to 92\%. However, incorporating multiple input modalities simultaneously or applying one-shot prompts does not consistently enhance performance and may degrade results. Furthermore, higher temperature values reduce performance. Based on these results, we recommend using screenshot inputs with zero temperature to maximize accuracy for LLM-based detectors with HTML serving as auxiliary context when screenshot information is insufficient.
\end{abstract}


\section{Introduction} \label{sec:introduction}

Phishing attacks have become a pervasive global threat via fraudulent websites that impersonate legitimate services (\eg, facebook.com)~\cite{FBIRelea95:online}. 
In the ongoing battles against phishing attacks, traditional detection approaches examine URLs~\cite{ghalechyan2024phishing,Rani2024PhishingURL,Prasad2024PhiUSIIL}, HTML~\cite{Opara2020HTMLPhish,Purwanto2022PhishSim,Subramani2022PhishInPatterns}, logos~\cite{Lin2021Phishpedia,Kirlappos2012SP,bozkir2020logosense}, screenshots~\cite{Abdelnabi2020VisualPhishNet,Trinh2022Visual}, and the combinations of these features~\cite{van2021textvision,Liu2022PhishIntention,KnowPhish2024Usenix} to identify fraudulent websites.
However, these approaches are typically trained on a subset of popular brands, constraining their ability to detect less common or newly established websites. These approaches also lack interpretability, providing little explanation to users. This opacity hinders users' ability to understand the rationale behind phishing classifications, potentially reducing trust in detection systems and limiting the practical deployment in the wild~\cite{Charmet2024VORTEX}.

Recently, large language models (LLMs) have gained significant attention for phishing detection~\cite{cao2025phishagent,Liu2024PhishLLM,Chatbots2024SP} because they are pre-trained on massive datasets and can
process diverse input modalities and offer interpretable explanations for their decisions~\cite{openai2024,gemini2025}.
However, prior work offers limited insights into how we effectively leverage LLMs for phishing detection. 
This gap reveals several key limitations in the current literature: (1) the absence of systematic evaluations assessing the detection performance across LLMs and deep learning-based methods; (2) limited evaluations of brand identification performance for explainability; (3) insufficient assessment of the individual and combined effectiveness of different input components (URLs, HTML, logos, screenshots, and their combinations) in LLM-based phishing brand identification systems; (4) lack of systematic investigations of how model settings and parameters influence LLM-based brand identification performance; and (5) inadequate in-depth analysis of failure cases and their potential causes.

To address these limitations, we pose four research questions to guide effective use of LLMs for phishing detection:
\begin{itemize}[leftmargin=*, itemsep=2pt, parsep=0pt, topsep=0pt, partopsep=0pt]
    \item \textbf{RQ1:} How effectively can LLMs detect phishing websites using their inherent knowledge and reasoning capabilities, and do they outperform traditional deep learning models?

    \item \textbf{RQ2:} What are the underlying failure causes of LLMs in phishing detection?

    \item \textbf{RQ3:} How do \textit{individual} input modalities and their \textit{combinations} affect the performance of LLMs in phishing brand identification?

    \item \textbf{RQ4:} What is the impact of parameter configurations and prompt design on the effectiveness and reliability of LLMs in phishing brand identification? 

\end{itemize}

To evaluate LLMs and traditional DL models (RQ1 and RQ2), we first design URL-Screenshot-HTML inputs using 19,131 real-world phishing websites and analyze LLM-generated responses by \texttt{Gemini}. We find that commercial LLMs perform best, with \texttt{GPT} reaching the highest detection rate of 93.86\% on phishing data. Deep learning (DL) models are fast and accurate on benign data ($\geq$97\%), but ineffective on phishing data due to high false negative rates ($\geq$64\%). Our in-depth analysis shows that LLM false negatives mainly arise from missing phishing signals, while false positives stem from sensitivity to data requests.

For RQ3 and RQ4, we design prompts based on \textit{individual} and \textit{multimodal} components with temperature variation and evaluate component-wise performance across popular LLMs.
Results show that most models perform better at lower temperatures with screenshot-related inputs.
Particularly, \texttt{Gemini} (94.59\%) is a promising LLM for brand identification with high accuracy and low cost,
while \texttt{Qwen} (91.97\%) stands out among open-source LLMs for its comparable performance and moderate latency.
%
%
Word distribution analysis shows that while prompts and multimodal inputs help LLMs generate relevant content, textual information from URLs and HTML can mislead predictions. Failures associated with screenshots and HTML inputs are often related to textual content, whereas logo-related failures are primarily linked to color and style. 


This study makes the following key contributions:

\begin{itemize}[leftmargin=*, itemsep=2pt, parsep=0pt, topsep=0pt, partopsep=0pt]
    \item We conduct a comprehensive study of popular commercial and open-source LLMs alongside deep learning-based models for phishing detection using 19,131 real-world phishing websites. Results show that commercial LLMs achieve high true positive rates, while deep learning models excel in true negative detection.
    \item We study the impact of input components and model settings on brand identification across popular commercial and open-source LLMs, and identify optimal configurations.
    \item We analyze LLM responses (word distributions of failure cases) to uncover potential tendencies. Our findings suggest that detection failures are related to phishing signals and content mismatch, identification failures are input-dependent, with screenshot and HTML inputs often linked to textual confusion, and logo-related failures tied to color and style features.
    \item Based on our findings, we recommend using screenshot-based inputs as the primary input for LLM-based phishing brand identification at low temperatures (\ie, 0.0), with HTML as a complementary input for post-processing.
    \item We will publicly share our collected dataset.
\end{itemize}

\section{Background} \label{sec:background}

\subsection{Phishing Attack and Defense}
\PP{Phishing Attack}Phishing is a type of social engineering attack where fraudulent websites resemble the appearance of legitimate websites to trick victims into disclosing their sensitive information, which is misused for further fraud or other criminal activities.

\PP{Phishing Detectors}Phishing detectors aim to detect malicious websites and prevent users from accessing them.
Recent deep learning-based phishing detection systems~\cite{Lin2021Phishpedia,Liu2022PhishIntention} leverage multiple sources of information, such as HTML content, visual appearance, and other metadata (\eg, domain age, TLS certificate information), to improve detection accuracy. 
These systems rely on predefined reference lists to compare suspected websites with legitimate ones for phishing detection and to infer the intended targets of the webpage.

\subsection{Large Language Models (LLMs)}
\PP{LLMs}
LLMs are AI systems that can understand and generate human-like text. 
LLMs can be categorized into commercial LLMs (\eg, \texttt{ChatGPT}~\cite{GPT:online}), which offer APIs but are limited by cost and access, and open-source LLMs (\eg, \texttt{Qwen}~\cite{Qwen:online}), which are more accessible and customizable.

\PP{LLM Configuration}
Configuration refers to parameters and settings that control how LLMs behave during inference. \textit{Temperature} is a critical parameter as it controls output randomness and creativity. 
In phishing detection contexts, its impact remains underexplored. Theoretically, low temperatures \textit{would} produce deterministic, stable classifications, while high temperatures \textit{may} introduce randomness and inconsistency. However, 
systematic empirical investigations validating these assumptions have been little explored.

\PP{Prompt Engineering}Prompt engineering designs inputs to elicit desired outputs from LLMs, where input structure dramatically affects output quality.
In \textit{zero-shot prompting}~\cite{gpt3}, LLMs rely solely on their pre-trained knowledge to generate responses.
\textit{One-shot prompting}~\cite{gpt3} provides LLMs with a single example to guide responses. This example serves as a guide for models in answering similar questions for other inputs.
\aref{sec:oneshot_example} illustrates a one-shot example where a \textit{Chase} logo guides brand identification.

\PP{LLM-based Phishing Detectors}
The advances of LLMs have enabled the application for phishing detection tasks~\cite{Liu2024PhishLLM,lee2024mmlm,koide2024ChatPhishDetector}. The general pipeline for LLM-based phishing detectors involves several steps. First, cleaning the dataset and ensuring that the input tokens comply with the token limitations of the models. Prompts are then constructed from various resources and fed into the LLMs, leveraging their internal reasoning capabilities to perform the detection. Post-processing is applied to extract relevant outputs and determine whether a given input is phishing. 


\section{Evaluation Design} \label{sec:eval_design}
\begin{figure}[!t]
\centering
\includegraphics[width=0.98\columnwidth]{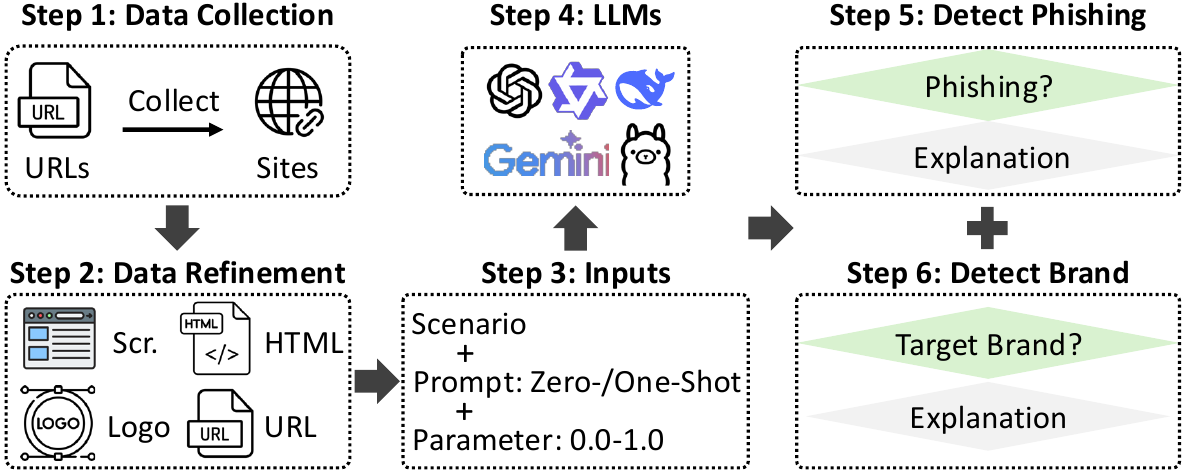}
\caption{\textbf{Overview of Our Evaluation Experiment.}}

~\label{fig:overview}
\end{figure}

\subsection{Overview of Our Evaluation}

As illustrated in~\autoref{fig:overview}, our methodology evaluates LLMs for phishing detection and brand identification through six key steps:
(1) collecting phishing (URLs, HTML, and screenshots) and benign data (\autoref{subsec:data}), (2) refining the dataset via clustering, sampling, and cleaning (\autoref{subsec:data}), (3) constructing inputs by combining scenarios, prompts, and parameters (\autoref{subsec:setup}), (4) select representative LLMs (\autoref{subsec:model}), (5) detecting phishing with explanations (\autoref{subsec:eval_plan}), and (6) identifying target brands (\autoref{subsec:eval_plan}).

\subsection{Data Collection and Refinement} \label{subsec:data}

\PP{Web Crawler Design}
We begin by collecting a comprehensive real-world phishing dataset comprising URLs, HTML content, and webpage screenshots. 
The primary source is the APWG eCrime Exchange (eCX)~\cite{APWGTheA96:online}, which represents one of the largest repositories of verified real-world phishing websites and has been extensively used in previous work~\cite{Ji2025Usenix,Zhang2022SPARTACUS,oest2019phishfarm,oest2020sunrise,Lim2025What}. 
As APWG eCX provides only phishing URLs, we develop a web crawler that, at one-minute intervals, collects (1) client-side resources (\eg, HTML content) and (2) screenshots. 
The crawler is implemented using \texttt{Google Selenium Chrome WebDriver}~\cite{Selenium:online} to simulate authentic user interactions, ensuring complete loading and rendering of all client-side resources. 
This approach also helps circumvent basic anti-bot detection mechanisms commonly deployed by phishing websites~\cite{amin2020web,li2021good}.
Furthermore, we use the same crawler to obtain benign websites.

\begin{table}[!t]
    \centering
    \footnotesize
    \caption{\textbf{Overview of Datasets (Jan. - Sep. 2024, 9 mon).}} 
    \label{tab:vision_statistics}
    \begin{tabular}{lrr}
        \toprule
        \multicolumn{1}{c}{\textbf{Type}} & \multicolumn{1}{r}{\# \textbf{Brand}} & \multicolumn{1}{r}{\# \textbf{Sample}} \\
        \midrule
        Total APWG Reports         & ---   & 445,011                  \\
        Assessable Phishing URLs   & 1,628 & 296,537                  \\ \midrule \midrule
        Our Refined Final Dataset  & 253   & 19,131                   \\ \midrule 
        Logo Only                  & 197   & 8,117 (42.4\%)           \\  
        Login Form Only            & 29    & 625 \phantom{0}(3.3\%)   \\
        Logo \& Login Form         & 149   & 8,897 (46.5\%)\\
        None of Logo and Login Form & 66   & 1,492 \phantom{0}(7.8\%) \\ \midrule \midrule
        Benign Dataset & 243 & 243 \\ 
        \bottomrule
    \end{tabular}
\end{table}

\PP{Refining Collected Dataset}The overview of datasets is shown in~\autoref{tab:vision_statistics}.
APWG eCX provides 445,011 real-world phishing URLs from January 2024 to September 2024 (9 months). Our crawler successfully accesses 296,537 (66.64\%) phishing websites, while 33.36\% of inaccessible
websites are due to server shutdowns or network errors (\eg, DNS resolving errors). Among accessible samples, we use the provided brand labels and focus on the top 200 brands, covering 290,662 samples (98.02\% of 296.5k).
%
To ensure dataset quality and label accuracy, we use \texttt{Fastdup}~\cite{visualla87:online} to cluster screenshots, which is an open-source clustering tool that computes edge distances within graph components. 
We retain up to three representative samples per cluster (all if fewer than three). Subsequently, we conduct a two-person manual verification to validate brand labels and remove error pages.
The final dataset comprises 19,131 samples, 6.45\% out of 296,537, that span 253 brands. 
Note that the final brand number exceeds the initial 200 due to re-labeling samples whose original annotations do not match actual phishing visual content.
%

Original reference lists include only 53 brands present in the phishing dataset. To address this limitation, we expand the lists by adding 190 additional brands from existing websites, noting that 10 brands in the phishing dataset are no longer active. Therefore, the benign dataset covers 243 brands. 

\PP{Logo Image Collection from Screenshot}
To investigate the role of logos for LLMs in phishing detection, we use Faster-RCNN~\cite{Ren2017Fasterrcnn}, as implemented in Phishpedia~\cite{Lin2021Phishpedia}, to crop the logo regions from the screenshots. For each image, we extract up to the top three logo candidates based on detection confidence (all if fewer than three). As a result, 18,695 out of 19,131 screenshots (97.72\%) contain at least one detected logo. A sample is considered correctly matched if any detected logos correspond to the verified brand.

\PP{Tokenizing URLs for LLMs}
For the URLs in the dataset, token counts range from 6 to 4,965 when tokenized using the `Meta-Llama-3-8B' tokenizer~\cite{metallam96:online}, with only one URL exceeding 4,096 tokens. This indicates that the vast majority of URLs are relatively short and well within the input capacity of LLMs, ensuring efficient processing and minimizing the risk of truncation during inference.

\PP{Tokenizing HTML Contents for LLMs}HTML content is often long and exceeds context limits. To address this, we analyze the HTML structure across 19,131 samples, identify 1,442 unique tags, and select the top 200 most frequent tags (appear $\geq$ 30 times). 
Additionally, we remove tags related to external links, styling, scripts, CSS, and other non-informative elements~\cite{aljofey2022effective}.
After preprocessing, token counts range from 14 to 114,921, with 95.91\% of samples under 4,096 tokens and all samples below 131,072 tokens. 
For the small subset that exceeds the maximum token limit, we extract and truncate the textual content to ensure compatibility with the models.


\subsection{LLM Selection} \label{subsec:model}

We carefully select 7 LLMs for our evaluation, as summarized in \aref{sec:appendix_llm_infor}-\autoref{tab:llm_info_sum}. For clarity, we refer to the models by their names as listed in the table throughout the paper. Specifically, we include two commercial models, \texttt{Gemini}~\cite{Gemini:online} (Google) and \texttt{GPT}~\cite{GPT:online}  (OpenAI). \texttt{Gemini} is hosted on Google Cloud~\cite{GoogleCloud:online} while \texttt{GPT} is hosted by Microsoft Azure~\cite{MicrosoftAzure:online}. Also, we incorporate several recently released open-source vision-language models, such as \texttt{Janus}~\cite{Janus:online}, \texttt{VL2}~\cite{VL2:online}, \texttt{Llama}~\cite{Llama:online}, and \texttt{Qwen}~\cite{Qwen:online}.
To provide a comparative baseline, we employ the R1~\cite{R1Distill:online} text-based LLM. These open-source models are deployed on NVIDIA A30 GPUs, except for VL2, which runs on A6000 GPUs.

\subsection{Input Construction for LLM} \label{subsec:setup}


\PP{Prompt Engineering}
For phishing detection, we design prompts by combining scenarios, tasks, and related inputs. For phishing brand identification, we follow the prompt structure established by Lee et al.~\cite{lee2024mmlm} and design inputs with three components: task information, answer instructions, and optional examples. 
Zero-shot inputs incorporate only the task and answer instructions, while one-shot inputs add an example to guide model behavior.
\autoref{tab:prompt_example} shows a one-shot logo-based input. The task information lets LLMs identify brands. The instruction then specifies the expected response format with brands and evidence. A \textit{Chase} logo with its expected response is provided as guidance, followed by a new logo for answering.
Further details are available in our open-science release.


\begin{itemize}[leftmargin=*, itemsep=2pt, parsep=0pt, topsep=2pt, partopsep=0pt]
    \item \textit{Task Information}: A brief description of the inference task to establish context.
    \item \textit{Answer Instruction}: A clear directive guiding the LLMs on how to generate the response.
    \item \textit{One-Shot Example}: A single example demonstrating the expected input-output format based on the instruction.
\end{itemize}

\PP{Temperature Configuration}We vary the temperatures (controlling output randomness) across five values: 0.0 or 0.1, 0.3, 0.5, 0.7, and 1.0. 
We exclude higher temperature values above 1.0 as LLMs tend to produce very diverse outputs, making a consistent comparison challenging. 
Note that when \textit{do\_sample} is set to \texttt{False}, we use the minimum temperatures, where 0.1 for \texttt{Qwen} and 0.0 for others. \texttt{GPT} is evaluated only at 0.0 owing to cost constraints.

%

\subsection{Evaluation and Analysis} \label{subsec:eval_plan}

After constructing inputs from various sources and parameters, we feed them into LLMs to detect phishing and identify target brands. Detection accuracy is measured against the website source (\autoref{sec:detection_results}), while brand recognition is evaluated via string matching with ground-truth annotations (\autoref{sec:identification_results}). 
We use brand identification to validate explainability in LLMs, as the brand is the most critical factor for interpretation and has been widely adopted in prior work~\cite{Liu2022PhishIntention,Dynaphish2023USenix,Ji2025Usenix,lee2023attacking,Hao2024Diffusion}.
To understand why LLMs may fail, we analyze responses for both tasks to uncover the specific features they emphasize. 


\PP{Detection}We randomly select up to 100 misclassified phishing samples per predicted category of LLMs (or all if fewer) and consider all misclassified benign samples. We then use \texttt{Gemini} to summarize the underlying reasons from detection explanations, categorize the reasons into groups, and report statistics. For DL-based models, failure statistics are computed directly from the misclassified samples. More details are provided in~\autoref{subsec:detection_fail}.

\PP{Brand identification}
We analyze word distributions in LLM responses to uncover the semantic cues emphasized during prediction.
We use spaCy~\cite{spaCy:online} instead of \texttt{Gemini} to avoid potential bias from LLMs for interpretation, since brand identification requires more fine-grained analysis than detection.
We begin by extracting words and their linguistic features, then remove stop words (\eg, \textit{for}, \textit{the}), irrelevant terms, and brand names (\eg, \textit{Facebook}). We exclude brand names because the goal of LLMs is to recognize them; removing these names helps prevent frequency bias and ensures that the analysis reflects the semantic and structural cues the models rely on. 
We also exclude terms (\eg, \textit{screenshot}) that are mentioned in the prompts and serve as clear indicators for brand identification.
This process yields 264 words, which are grouped into six categories, as shown in~\autoref{tab:explanation_category}.
More details are provided in~\autoref{sec:analysis}.

\section{Phishing Detection Performance} \label{sec:detection_results}
\begin{table*}[!t]
    \centering
    \caption{\textbf{Phishing Detection Performance With URL-HTML-Screenshot Inputs.} Bold values denote the best performance.}
    \label{tab:result_llm_dl}
    \resizebox{\linewidth}{!}{
    \begin{tabular}{l l r r r r r r r r r r r}
        \toprule
        \multirow{2}{*}{\textbf{Dataset}} & \multirow{2}{*}{\textbf{Type}} & \multirow{2}{*}{\textbf{Model}} & \multicolumn{4}{c}{\textbf{Predict Results}} & \multicolumn{3}{c}{\textbf{Inference Time (Sec.)}} & \multicolumn{3}{c}{\textbf{Response Length$^*$}} \\ \cmidrule{4-7} \cmidrule{8-10} \cmidrule{11-13}
        & & & Phishing & Benign & Uncertain & Block & Min & Max & Mean & Min & Max & Mean \\\midrule
        \multirow{10}{*}{\rotatebox{90}{Phishing (19,131)}} 
        & \multirow{6}{*}{\raggedright\makecell[l]{\textbf{LLMs}}} 
          & GPT    & \textbf{17,957 (93.86\%)} &  181 \phantom{0}(0.95\%) &    993 \phantom{0}(5.19\%) & 0 (0.00\%)   & 2.05 & 605.32 & 5.32 & 35 & 99  & 62.82\\
        & & Gemini & 17,308 (90.47\%)          &  544 \phantom{0}(2.84\%) &  1,279 \phantom{0}(6.69\%) & 0 (0.00\%)   & 1.48 & 80.82  & 2.68 & 13 & 89  & 39.19\\
        & & Qwen   & 12,476 (65.21\%)          &  995 \phantom{0}(5.20\%) &            5,660 (29.59\%) & 0 (0.00\%)   & 3.23 & 16.45  & 5.88 & 21 & 165 & 70.39\\
        & & Llama  & 13,403 (70.06\%)          &  5,490 (28.70\%)         &     42 \phantom{0}(0.22\%) & 196 (1.02\%) & 1.13 & 14.09  & 4.29 & 1  & 168 & 31.10\\
        & & Janus  &  2,650 (13.09\%)          &  2,505 (13.85\%)         &     13,976 (73.05\%)       & 0 (0.00\%)   & 0.01 & 32.83  & 2.65 & 11 & 391 & 37.71\\
        & & VL2    &  9,034 (47.22\%)          & 10,044 (52.50\%)         &     53 \phantom{0}(0.28\%) & 0 (0.00\%)   & 0.59 & 5.92   & \textbf{2.20} & 2  & 67  & 24.73\\\cmidrule{2-13}
        & \multirow{4}{*}{\raggedright\makecell[l]{\textbf{DL}}} 
        & PhishIntention                    & 5,340 (27.91\%)          & 13,749 (71.87\%) & 42 (0.22\%) &  0 (0.00\%) & NA & NA &               1.20 & NA & NA & NA$^{\dagger}$\\
        & & PhishIntention-Ref$^{\ddagger}$ & 6,089 (31.83\%)          & 13,000 (67.95\%) & 42 (0.22\%) &  0 (0.00\%) & NA & NA &               1.37 & NA & NA & NA\\ 
        & & Phishpedia                      & 6,739 (35.23\%)          & 12,350 (64.55\%) & 42 (0.22\%) &  0 (0.00\%) & NA & NA & \textbf{1.07}      & NA & NA & NA\\ 
        & & Phishpedia-Ref                  & \textbf{8,259 (43.17\%)} & 10,830 (56.61\%) & 42 (0.22\%) &  0 (0.00\%) & NA & NA & 1.25               & NA & NA & NA\\\midrule \midrule
        \multirow{10}{*}{\rotatebox{90}{Benign (243)}} 
        & \multirow{6}{*}{\raggedright\makecell[l]{\textbf{LLMs}}} 
          & GPT    &  7 \phantom{0}(2.88\%) & 182 (74.90\%) &  54 (22.22\%)           & 0 (0.00\%) & 2.69 & 18.30 & 5.29 & 45 & 86  & 61.61\\
        & & Gemini & 15 \phantom{0}(6.17\%) & 217 (89.30\%) &  11 \phantom{0}(4.53\%) & 0 (0.00\%) & 1.96 & 5.48  & 2.87 & 18 & 58  & 32.50\\
        & & Qwen   &  3 \phantom{0}(1.23\%) & 165 (67.90\%) &  75 (30.86\%)           & 0 (0.00\%) & 3.88 & 11.93 & 5.54 & 44 & 106 & 64.46\\
        & & Llama  & 16 \phantom{0}(6.58\%) & \textbf{227 (93.42\%)} &   0 \phantom{0}(0.00\%) & 0 (0.00\%) & 3.63 & 12.82 & 5.01 & 13 &  86 & 32.59\\
        & & Janus  &  2 \phantom{0}(0.82\%) & 118 (48.56\%) & 123 (50.62\%)           & 0 (0.00\%) & 1.36 & 7.11  & 2.84 & 18 &  70 & 37.95\\
        & & VL2    & 96 (39.51\%)           & 42 (17.28\%)  & 105 (43.21\%)           & 0 (0.00\%) & 1.53 & 6.16  & \textbf{2.66} & 14 &  78 & 29.35\\\cmidrule{2-13}
        & \multirow{4}{*}{\raggedright\makecell[l]{\textbf{DL}}} 
        & PhishIntention       & 1 (0.41\%) & \textbf{242 (99.59\%)} & 0 (0.00\%) & 0 (0.00\%) & NA & NA & 1.48 & NA   & NA & NA\\
        & & PhishIntention-Ref & 1 (0.41\%) & \textbf{242 (99.59\%)} & 0 (0.00\%) & 0 (0.00\%) & NA & NA & 3.26 & NA   & NA & NA\\
        & & Phishpedia         & 6 (2.47\%) & 237 (97.53\%) & 0 (0.00\%) & 0 (0.00\%) & NA & NA & 1.26 & NA   & NA & NA\\ 
        & & Phishpedia-Ref     & 3 (1.23\%) & 240 (98.77\%) & 0 (0.00\%) & 0 (0.00\%) & NA & NA & \textbf{1.24} & NA   & NA & NA\\ \bottomrule
        \multicolumn{13}{l}{$^*$Response Length represents the average length of model-generated replies, measured in words after text cleaning.}\\
        \multicolumn{13}{l}{$^{\dagger}$NA indicates that the model is not applicable to this scenario.} \\
        \multicolumn{13}{l}{$^{\ddagger}$Ref indicates that the model incorporates additional references for the 190 brands.}
    \end{tabular}
    }
\end{table*}

\begin{table*}[!t]
    \centering
    \caption{\textbf{Detection Failure Statistics by Reason Category.}}
    \label{tab:result_detection_fail}
    \footnotesize
    \resizebox{0.95\linewidth}{!}{
    \begin{tabular}{l l r r r r r r r r r}
        \toprule
        \multirow{3}{*}{\textbf{Dataset}} & \multirow{3}{*}{\textbf{Misclassify}} & \multirow{3}{*}{\textbf{Model}} & \multirow{3}{*}{\textbf{Sample}} & \multicolumn{7}{c}{\textbf{Reason Category}} \\ \cmidrule{5-11}
        & & & & \multirow{2}{*}{\makecell{Data \\Request}} & \multirow{2}{*}{Insufficient} & \multirow{2}{*}{Mismatch} & \multirow{2}{*}{\makecell{No Phishing\\ Signals}} & \multirow{2}{*}{\makecell{No Trust\\ Signals}} & \multirow{2}{*}{\makecell{Suspicious\\ URLs}} & \multirow{2}{*}{Trusted} \\
        & \\ \midrule
        \multirow{12}{*}{\rotatebox{90}{Phishing}} 
        & \multirow{6}{*}{\raggedright\makecell[l]{\textbf{Benign}}} 
        &   GPT    & 100 & 0 \phantom{0}(0.0\%) & 0 \phantom{0}(0.0\%) & 3 (3.0\%) & 100 (100.0\%)          & 2 (2.0\%) & 3 (3.0\%) & 63 (63.0\%) \\
        & & Gemini & 100 & 2 \phantom{0}(2.0\%) & 1 \phantom{0}(1.0\%) & 3 (3.0\%) & 100 (100.0\%)          & 0 (0.0\%) & 1 (1.0\%) & 76 (76.0\%) \\
        & & Qwen   & 100 & 0 \phantom{0}(0.0\%) & 0 \phantom{0}(0.0\%) & 0 (0.0\%) & 0 \phantom{00}(0.0\%)  & 0 (0.0\%) & 2 (2.0\%) & 96 (96.0\%) \\
        & & Llama  & 100 & 5 \phantom{0}(5.0\%) & 5 \phantom{0}(5.0\%) & 7 (7.0\%) & 95 \phantom{0}(95.0\%) & 0 (0.0\%) & 3 (3.0\%) & 83 (83.0\%) \\
        & & Janus  & 100 & 0 \phantom{0}(0.0\%) & 0 \phantom{0}(0.0\%) & 3 (3.0\%) & 98 \phantom{0}(98.0\%) & 0 (0.0\%) & 0 (0.0\%) & 96 (96.0\%) \\
        & & VL2    & 100 & 33 (33.0\%)          & 18 (18.0\%)          & 2 (2.0\%) & 79 \phantom{0}(79.0\%) & 0 (0.0\%) & 1 (1.0\%) & 28 (28.0\%) \\\cmidrule{2-11}
        & \multirow{6}{*}{\raggedright\makecell[l]{\textbf{Uncertain}}} 
        &   GPT    & 100 & 11 (11.0\%)          & 93 \phantom{0}(93.0\%) & 44 (44.0\%)          & 13 (13.0\%)          & 26 (26.0\%)          & 61 (61.0\%)          & 0 (0.0\%) \\
        & & Gemini & 100 & 6 \phantom{0}(6.0\%) & 100 (100.0\%)          & 13 (13.0\%)          & 4 \phantom{0}(4.0\%) & 4 \phantom{0}(4.0\%) & 54 (54.0\%)          & 3 (3.0\%) \\
        & & Qwen   & 100 & 23 (23.0\%)          & 100 (100.0\%)          & 30 (30.0\%)          & 4 \phantom{0}(4.0\%) & 10 (10.0\%)          & 37 (37.0\%)          & 2 (2.0\%) \\
        & & Llama  & 42  & 2 \phantom{0}(4.8\%) & 41 \phantom{0}(97.6\%) & 0 \phantom{0}(0.0\%) & 5 (11.9\%)           & 3 \phantom{0}(7.1\%) & 4 \phantom{0}(9.5\%) & 0 (0.0\%) \\
        & & Janus  & 100 & 40 (40.0\%)          & 63 \phantom{0}(63.0\%) & 38 (38.0\%)          & 19 (19.0\%)          & 40 (40.0\%)          & 44 (44.0\%)          & 8 (8.0\%) \\
        & & VL2    & 53  & 30 (56.6\%)          & 30 \phantom{0}(56.6\%) & 3 \phantom{0}(5.7\%) & 12 (22.6\%)          & 4 \phantom{0}(7.5\%) & 4 \phantom{0}(7.5\%) & 4 (7.5\%) \\\midrule \midrule        
        \multirow{12}{*}{\rotatebox{90}{Benign}} 
        & \multirow{6}{*}{\raggedright\makecell[l]{\textbf{Phishing}}} 
          & GPT    & 7  & 7 (100.0\%)            & 0 \phantom{0}(0.0\%) & 7 (100.0\%)            & 0 \phantom{0}(0.0\%) & 1 \phantom{0}(14.3\%) & 7 (100.0\%)            & 0 (0.0\%) \\
        & & Gemini & 15 & 15 (100.0\%)           & 0 \phantom{0}(0.0\%) & 14 \phantom{0}(93.3\%) & 0 \phantom{0}(0.0\%) & 0 \phantom{00}(0.0\%) & 12 \phantom{0}(80.0\%) & 1 (6.7\%) \\
        & & Qwen   & 3  & 3 (100.0\%)            & 0 \phantom{0}(0.0\%) & 3 (100.0\%)            & 0 \phantom{0}(0.0\%) & 0 \phantom{00}(0.0\%) & 3 (100.0\%)            & 0 (0.0\%) \\
        & & Llama  & 16 & 14 \phantom{0}(87.5\%) & 1 \phantom{0}(6.3\%) & 12 \phantom{0}(75.0\%) & 0 \phantom{0}(0.0\%) & 4 \phantom{0}(25.0\%) & 13 \phantom{0}(81.3\%) & 1 (6.3\%) \\
        & & Janus  & 2  & 0 \phantom{00}(0.0\%)  & 0 \phantom{0}(0.0\%) & 2 (100.0\%)            & 0 \phantom{0}(0.0\%) & 2 (100.0\%)           & 2 (100.0\%)            & 0 (0.0\%) \\
        & & VL2    & 96 & 66 \phantom{0}(68.8\%) & 32 (33.3\%) & 41  \phantom{0}(42.7\%)         & 11 (11.5\%)          & 1 \phantom{00}(1.0\%) & 14 \phantom{0}(14.6\%) & 6 (6.3\%) \\ \cmidrule{2-11}
        & \multirow{6}{*}{\raggedright\makecell[l]{\textbf{Uncertain}}} 
        &     GPT  & 54  & 7 (13.0\%)           & 54 (100.0\%)            & 43 (79.6\%)           & 3 \phantom{0}(5.6\%) & 8 (14.8\%) & 1 \phantom{0}(1.9\%) & 8 (14.8\%)           \\
        & & Gemini & 11  & 0 \phantom{0}(0.0\%) & 11 (100.0\%)            & 1 \phantom{0}(9.1\%)  & 1 \phantom{0}(9.1\%) & 0 \phantom{0}(0.0\%)  & 0 \phantom{0}(0.0\%) & 0 \phantom{0}(0.0\%) \\
        & & Qwen   & 75  & 15 (20.0\%)          & 75 (100.0\%)            & 4 \phantom{0}(5.3\%)  & 18 (24.0\%)          & 6 \phantom{0}(8.0\%)  & 3 \phantom{0}(4.0\%) & 11 (14.7\%)          \\
        & & Llama  & 0   & 0 \phantom{0}(0.0\%) & 0 \phantom{00}(0.0\%)   & 0 \phantom{0}(0.0\%)  & 0 \phantom{0}(0.0\%) & 0 \phantom{0}(0.0\%)  & 0 \phantom{0}(0.0\%) & 0 \phantom{0}(0.0\%) \\
        & & Janus  & 123 & 13 (10.6\%)          & 102 \phantom{0}(82.9\%) & 12 \phantom{0}(9.8\%) & 49 (39.8\%)          & 63 (51.2\%)           & 23 (18.7\%)          & 25 (20.3\%)          \\
        & & VL2    & 105 & 13 (12.4\%)          & 49 \phantom{0}(46.7\%)  & 6 \phantom{0}(5.7\%)  & 47 (44.8\%)          & 0 \phantom{0}(0.0\%)  & 3 \phantom{0}(2.9\%) & 35 (33.3\%)          \\ \bottomrule
    \end{tabular}
    }
\end{table*}

We evaluate the phishing detection accuracy of six LLMs (\texttt{GPT}, \texttt{Gemini}, \texttt{Qwen}, \texttt{Llama}, \texttt{Janus}, and \texttt{VL2}) with URLs, screenshots, and HTML as inputs at the minimum temperature setting, along with two recent deep learning–based models (\texttt{PhishIntention~\cite{Liu2022PhishIntention}} and \texttt{Phishpedia}~\cite{Lin2021Phishpedia}) as baselines.
We begin by presenting the overall phishing detection performance in~\autoref{subsec:detection_general}, followed by an in-depth analysis of misclassifications in \autoref{subsec:detection_fail}.

\subsection{Overall Detection Performance}
\label{subsec:detection_general}
\autoref{tab:result_llm_dl} shows phishing detection performance on both phishing and benign datasets. Commercial LLMs achieve the highest detection rates on the phishing dataset. Specifically, \texttt{GPT} and \texttt{Gemini} successfully detect 93.86\% and 90.47\% of phishing websites, with low false negative rates (phishing misclassified as benign) of 0.95\% and 2.84\%, respectively. On the benign dataset, however, both models demonstrate non-negligible false positive rates (2.88\% for \texttt{GPT} and 6.17\% for \texttt{Gemini}), and \texttt{GPT} exhibits higher uncertainty (22.22\%). \texttt{Gemini} is faster than \texttt{GPT} (2.87s vs. 5.29s per query), whereas \texttt{GPT} produces longer and more detailed outputs (average 61-63 words) compared to \texttt{Gemini} (32-40 words).

Performance among open-source LLMs varies considerably. \texttt{Qwen} detects 65.21\% of phishing websites, with a moderate false negative rate of 5.20\%, but a high level of uncertainty (29.59\%). On the benign dataset, it produces only 1.23\% of false positive rate, yet again exhibits a large proportion of uncertain predictions (30.86\%), reflecting its overly cautious behavior.
\texttt{Llama} performs comparatively well, detecting 70.06\% of phishing websites, but misclassifies 28.70\% as benign and shows a very low uncertainty (0.22\%). Notably, \texttt{Llama} exhibits a 1.02\% block rate, returning the response `I cannot assist you with that.'
On the benign dataset, it achieves a high true negative rate of 93.42\%. However, its efficiency is limited by an average inference time of 5.01 seconds, which is slower than most other models.
In contrast, \texttt{Janus} demonstrates extremely poor performance, detecting only 13.09\% of phishing attacks while producing 73.05\% uncertain outputs. On the benign samples, it is uncertain about more than half (50.62\%) of the samples.
\texttt{VL2} also performs poorly, detecting only 47.22\% of phishing websites while misclassifying more than half (52.50\%) as benign. On the benign dataset, its accuracy drops sharply, with only 17.28\% of benign websites correctly identified, and 43.21\% of samples are uncertain. Although it has the fastest inference time among open-source LLMs (average 2.20-2.66 seconds), its low reliability makes it unsuitable for practical deployment.

\PP{Deep Learning-based Detectors}
The deep learning-based models all perform poorly on phishing detection due to high false negative rates. Incorporating reference lists improves their performance somewhat, but detection accuracy remains limited. Their inference time is relatively fast, averaging around 1.0-1.4 seconds. On the benign dataset, however, these models perform well, achieving 97.53-99.59\% accuracy.


\observ{Commercial LLMs achieve the best performance on the phishing dataset, while \texttt{Llama} performs comparatively better among open-source LLMs but occasionally refuses to respond. Deep learning-based models are fast and accurate on the benign data, but are largely ineffective due to high false negative rates}

\subsection{In-Depth Analysis of Misclassification} \label{subsec:detection_fail}  

We conduct an in-depth analysis of detection failures. 
For LLMs, we randomly select up to 100 misclassified samples (or all if fewer) from the phishing dataset in each category, and review all misclassified samples from the benign dataset. We then use \texttt{Gemini} to summarize the underlying reasons from the detection explanations, categorize the reasons into distinct groups in \autoref{tab:fail_explanation_category}, and present the corresponding statistics in \autoref{tab:result_detection_fail}. For DL-based models, failure statistics are computed directly from the misclassified samples.

\subsubsection{Misclassifications on Phishing Dataset}
\PP{Phishing as Benign (False Negative)}
\autoref{tab:result_detection_fail} shows that most LLMs, except for \texttt{Qwen}, primarily fail because they do not detect clear phishing signals. In such cases, the models are more likely to classify the website as benign. Another influential factor is the presence of trusted indicators, which further leads the LLMs to misclassifications.

\PP{Phishing as Uncertain}For samples misclassified as uncertain, the main factor is insufficient information for LLMs to decide. One primary cause is truncated HTML context due to context length limits, which removes some elements and creates mismatches. 
To further investigate the impact of incomplete HTML, we re-run the entire phishing and benign datasets using \texttt{Gemini} and analyze the effect of HTML information loss on model performance.

\PP{Phishing is Blocked}\texttt{Llama} blocks 196 samples by responding with messages such as `I cannot assist with that request.' Upon reviewing the screenshots, we find that 176 samples (90\%) belong to social media platforms. Additionally, 47 samples (24\%) are related to sexually explicit content, while 68 samples (35\%) contain only a large logo or a partial logo within the screenshot.

\PP{DL Misclassification}
As shown in \autoref{tab:result_llm_dl}, the augmented lists improve performance. \texttt{PhishIntention} rises from 27.91\% to 31.83\% on the phishing dataset while remaining stable on benign samples. \texttt{Phishpedia} increases from 35.23\% to 43.17\% on the phishing dataset and from 97.53\% to 98.77\% on the benign dataset. Additionally, 42 samples remain uncertain because of hitting forbidden words.

Another reason arises from model structures and dataset characteristics. As shown in \autoref{tab:vision_statistics}, 88.9\% of samples contain logos, but only 46.5\% contain both logos and login forms. As \texttt{Phishpedia} relies on logos for detection and \texttt{PhishIntention} additionally requires login forms, their maximum performance is inherently bounded by these proportions.

For the extended-reference models misclassified as benign, 95.58\% of failed cases for \texttt{Phishpedia} fall below the confidence threshold (0.87), while only 44 cases exceed the threshold. The remaining failures are due to domain-brand matching issues (\eg, top-level domain (TLD) or logo ratio inconsistencies). For \texttt{PhishIntention}, 54.45\% of failures fall below the threshold, while missing login forms or logos and TLD mismatches account for the remaining errors.


\observ{LLMs mainly produce false negatives due to missing phishing signals, and uncertain predictions from simplified HTML or content mismatches. \texttt{Llama} tends to block sexually explicit or social media samples, reflecting conservative behavior. Extending reference lists improves performance, while DL models misclassify largely due to dataset characteristics and inherent model biases}

\subsubsection{Misclassifications on Benign Dataset}

\PP{Benign as Phishing (False Positive)}
This category represents a critical type of false positive since it undermines usability and trust in detection systems. As shown in \autoref{tab:result_detection_fail}, most LLMs misclassify benign sites as phishing primarily due to the presence of data request elements and content mismatches (arising from simplified HTML structures or inconsistencies in information presentation). Additionally, some errors stem from domain or URL mismatches with legitimate ones, which trigger suspicion even when the site is authentic. 
These results reflect inherent limitations of LLMs, such as incomplete knowledge and insufficient contextual reasoning to differentiate legitimate service portals from phishing attempts. 
\looseness=-1

\PP{Benign as Uncertain}
\texttt{Janus} and \texttt{Qwen} achieve low false positives (0.82\% vs. 1.23\%) but high uncertainty (50.62\% vs. 30.86\%), making them unreliable for phishing detection. \texttt{Llama} shows no uncertainty, and \texttt{Gemini} (4.53\%) has a lower uncertainty rate compared to \texttt{GPT} (22.22\%). Most LLMs misclassify benign websites as uncertain due to insufficient information, often caused by simplified HTML structures or missing additional information such as SSL certificates.

\PP{DL Misclassification}
\texttt{PhishIntention} misclassifies one case because the screenshot contains multiple logos, and another failure case for the extended model due to the parent-child brand relationship. 
Similarly, \texttt{Phishpedia} misclassifies six samples due to missing login forms and an insufficient domain list. After extending the list, three additional failures are also caused by hierarchical brand relationships.

\observ{Misclassifications on the benign dataset largely stem from LLMs’ sensitivity to data requests, simplified HTML, and domain mismatches, reflecting limited brand knowledge. In contrast, DL-based models mainly fail due to hierarchical brand relationships or missing references, but achieve lower false positive rates when brand knowledge is included}

\section{Phishing Brand Identification Performance} \label{sec:identification_results}
We evaluate the brand identification accuracy of the seven LLMs (\texttt{GPT}, \texttt{Gemini}, \texttt{Qwen}, \texttt{Llama}, \texttt{Janus}, \texttt{VL2}, and \texttt{R1}) across various input prompts (\eg, prompts engineering techniques) and temperature settings. 
%
%
We first present the overall performance in~\autoref{subsec:general}.
Subsequently, in~\autoref{subsec:single}, we analyze how \textit{individual} modalities (including screenshots, logos, URLs, and HTML contents) affect performance across different temperature values -- the results are provided in~\autoref{tab:result_single}.
Based on the single-modality analysis, \autoref{subsec:combination} examines the impact of \textit{multimodal} input combinations (\autoref{tab:result_combination}) such as URLs with screenshots, URLs with logos, and URLs paired with both HTML content and screenshots.
\autoref{subsec:oneshot} explores the effect of one-shot prompting on model performance, providing insights into few-shot learning capabilities for phishing detection tasks. 

\subsection{Overall Performance} 
\label{subsec:general}

Commercial LLMs generally outperform open-source ones across all experiments and exhibit more stable behaviors across temperature settings.
With screenshot inputs, commercial LLMs achieve an average accuracy of 94.18\%, outperforming open-source models at 85.15\%.
A 9.03\% performance gap highlights the current advantage of commercial solutions in phishing brand identification tasks.

Among the commercial LLMs, \texttt{Gemini} outperforms \texttt{GPT} in recognizing brands from phishing websites while also incurring lower inference time and costs. \texttt{Gemini} achieves 94.59\% accuracy with screenshot and URL inputs, with an input cost of \$0.1 and output cost of \$0.7 per million tokens, and an average inference time of 1.62 seconds. In contrast, \texttt{GPT} incurs an input cost of \$2 and an output cost of \$8 per million tokens, with an average inference time of 3.64 seconds.
Meanwhile, among open-source LLMs, \texttt{Qwen} achieves performance comparable to that of commercial models under tasks that do not involve URLs. The best results are typically obtained at lower temperature values, with screenshot-related inputs except \texttt{Janus} and \texttt{R1}. Particularly, as shown in~\autoref{tab:result_single}, the best performance for each LLM is as follows: \texttt{GPT} with screenshot input at 93.95\% (temperature 0.0), \texttt{Gemini} with screenshot-URL input at 94.59\% (temperature 0.0), \texttt{Qwen} with screenshot input at 91.97\% (temperature 0.0), \texttt{Llama} with screenshot input at 89.26\% (temperature 0.0), \texttt{VL2} with screenshot input at 87.81\% (temperature 0.0). \texttt{Janus} achieves its highest accuracy with logo-URL input at 81.14\% (temperature 0.0), while the textual language model \texttt{R1} performs best with HTML input at 58.28\% (temperature 0.5).

All LLMs perform worse on textual inputs (\eg, URL-only, URL-One-shot, and HTML inputs) than on vision-related inputs for phishing brand identification. One-shot guidance slightly improves \texttt{Janus} but has little impact or negative impacts on other LLMs.

\observ{\texttt{Gemini} is a promising LLM for brand identification with screenshot-URL inputs at temperature 0.0 because of its lower cost and higher accuracy. \texttt{Qwen} also demonstrates strong potential as an open-source model, achieving competitive performance with screenshot inputs at a temperature of 0.1}

\begin{table}[!t]
    \centering
    \caption{\textbf{Brand Identification Performance Across Single Component Inputs and Temperatures.} Bold values denote the highest identification rate within each input category.}
    \label{tab:result_single}
    \resizebox{\linewidth}{!}{
    \begin{tabular}{p{0.09\linewidth} p{0.14\linewidth} rrrrr r}
        \toprule
        \multirow{2}{*}{\textbf{Input}} & \multirow{2}{*}{\textbf{Model}} & \multicolumn{5}{c}{\textbf{Temperature}} & \multirow{2}{*}{\textbf{Avg.}}\\
        \cmidrule(lr){3-7}
        & & 0.0/0.1 & 0.3 & 0.5 & 0.7 & 1.0 \\\midrule
        \multirow{6}{*}{\textbf{$^{**}$Scr.}} & GPT & 93.95 & $^*$NA & NA & NA & NA & 93.95\\
        & Gemini & \textbf{94.45} & 94.43 & 94.42 & 94.41 & 94.32 & 94.41\\
        & Qwen  & 91.97 & 90.84 & 91.02 & 90.89 & 90.32 & 91.01\\
        & Llama  & 89.26 & 88.41 & 88.01 & 86.88 & 85.80 & 87.67\\
        & Janus & 79.96 & 78.55 & 79.17 & 77.14 & 74.47 & 77.86\\
        & VL2 & 87.81 & 86.59 & 85.40 & 83.30 & 77.17 & 84.05\\\midrule
        \multirow{6}{*}{\textbf{Logo}} & GPT & 82.72 & NA & NA & NA & NA & 82.72\\
        & Gemini & 83.75 & 83.75 & \textbf{83.76} & 83.70 & 83.69 & 83.73\\
        & Qwen &  80.10 & 80.30 & 80.15 & 80.14 & 80.03 & 80.14\\
        & Llama & 79.30 & 79.22 & 78.98 & 78.97 & 78.76 & 79.05\\
        & Janus & 78.44 & 79.00 & 78.90 & 78.67 & 77.56 & 78.51\\
        & VL2 &  75.01 & 77.23 & 77.24 & 76.28 & 73.74 & 75.90\\\midrule
        \multirow{6}{*}{\textbf{URL}} & GPT & 28.81 & NA & NA & NA & NA & 28.81\\
        & Gemini &  \textbf{30.19} & 29.84 & 29.50 & 29.14 & 28.64 & 29.46\\
        & Qwen  & 14.00 & 13.91 & 14.06 & 14.43 & 14.27 & 14.13\\
        & Llama & 25.88 & 25.59 & 25.31 & 24.60 & 23.84 & 25.04\\
        & Janus & 24.38 & 24.04 & 23.49 & 22.85 & 21.26 & 23.20\\
        & VL2 & 20.32 & 19.92 & 19.18 & 18.89 & 17.71 & 19.20\\
        & R1 & 16.80  &  16.48 & 16.30  & 16.08 & 16.89 & 15.26\\\midrule
        \multirow{6}{*}{\textbf{HTML}} & GPT &  68.14 & NA & NA & NA & NA & 68.14\\
        & Gemini &  \textbf{71.88} & 71.78 & 71.70 & 71.56 & 71.48 & 71.68\\
        & Qwen &  65.27 & 65.14 & 64.97 & 64.69 & 64.08 & 64.83\\
        & Llama  & 60.33 & 60.03 & 59.97 & 59.46 & 58.71 & 59.70\\
        & Janus & 60.58 & 59.97 & 59.66 & 58.62 & 56.07 & 58.98\\
        & VL2 & 55.00 & 54.46 & 52.71 & 50.14 & 44.15 & 51.29\\
        & R1 & 57.86 & 58.10 & 58.28 & 57.79 & 56.48 & 57.70\\
        \bottomrule 
        \multicolumn{8}{l}{$^*$NA indicates that the model is either not applicable to this scenario or }  \\
        \multicolumn{8}{l}{the experiment was not conducted due to cost constraints.} \\
        \multicolumn{8}{l}{$^{**}$Scr. stands for screenshot.} \\
    \end{tabular}
    }
\end{table}

\subsection{Single Component with Temperature Conf.} \label{subsec:single}

We begin by evaluating how \textit{individual} input modalities affect phishing brand identification accuracy under different temperature configurations.
\autoref{tab:result_single} shows that screenshot inputs generally yield the best results, except for \texttt{Janus}, which performs better with logo inputs.

\subsubsection{Result of Screenshots}
\texttt{Gemini} achieves an average brand recognition rate of 94.41\% across various temperature settings, while \texttt{GPT} reaches 93.95\%. 
Among non-commercial models, \texttt{Qwen} performs best with an average brand detection rate of 91.01\%, followed by \texttt{Llama} (87.67\%), \texttt{VL2} (84.05\%), and \texttt{Janus} (77.86\%).

As the temperature increases, most LLMs exhibit performance declines, with minor fluctuations observed for \texttt{Qwen} and \texttt{Janus}. Specifically, \texttt{Gemini} remains stable with a maximum difference between the temperatures of 0.13\%, while performance differences for \texttt{Qwen}, \texttt{Llama}, \texttt{Janus}, and \texttt{VL2} are 1.65\%, 3.46\%, 5.49\%, and 10.64\%, respectively. These results demonstrate that \texttt{Gemini} is very stable for temperature change when using screenshot inputs, while \texttt{VL2} and \texttt{Janus} suggest greater sensitivity to randomness.

Overall, screenshot inputs are informative and effective for LLMs in recognizing brands. Although temperature impacts some models, lower temperature settings generally yield comparatively better predictions for screenshot-based input.

\subsubsection{Result of Logos}
The logo results in~\autoref{tab:result_single} show that \texttt{GPT} and \texttt{Gemini} also outperform non-commercial models on logo inputs. Particularly, \texttt{Gemini} achieves the highest average brand identification rate of 83.73\% and \texttt{GPT} reaches 82.72\%. \texttt{Qwen} performs best among non-commercial models with an average rate of 80.14\%, followed by \texttt{Llama} (79.05\%), \texttt{Janus} (78.51\%), and \texttt{VL2} (75.90\%).
Performance variation across temperatures is relatively small for most models, with performance differences of 0.07\%, 0.27\%, 0.54\%, and 1.44\% for \texttt{Gemini}, \texttt{Qwen}, \texttt{Llama}, and \texttt{Janus}, respectively. In contrast, \texttt{VL2} shows a larger fluctuation of 3.5\%.

\PP{Screenshot vs. Logo}
Logo inputs generally yield lower performance than screenshot inputs, with average performance drops of 10.68\% for \texttt{Gemini}, 11.23\% for \texttt{GPT}, 10.87\% for \texttt{Qwen}, 8.62\% for \texttt{Llama}, and 8.15\% for \texttt{VL2}. \texttt{Janus} is the only model that shows a slight improvement, with its performance increasing by 0.65\%.
In summary, commercial models still outperform non-commercial ones, but brand identification rates usually decrease with logos compared to screenshots. This suggests that screenshot inputs provide more comprehensive and effective information for LLMs in detecting brand information than logos alone.

\subsubsection{Results of URLs}
URL-only inputs in~\autoref{tab:result_single} yield poor results in both commercial and non-commercial models, with \texttt{Gemini} achieving the highest average brand recognition rate of 29.46\%.
%
As the temperature increases, most models drop in performance except \texttt{Qwen}, with an improvement from 14.00\% to 14.43\%.

\PP{Screenshots vs. Logos vs. URLs}
Compared to screenshot or logo inputs, LLM performance on URL inputs is much lower, as they typically contain irrelevant information or misleading characters that hinder LLMs from making correct predictions. However, if these URLs are used as the content for validating brand-URL consistency instead of direct inputs for LLMs, then the accuracy values can be interpreted as the potential error rates of such phishing detection models. This is because a small portion of URLs do contain brand-related information, and using LLMs to assess the consistency between predicted brands and URLs may result in errors.

\subsubsection{Results of HTML}
With HTML inputs, commercial models again demonstrate better performance. \texttt{Gemini} achieves the highest average brand detection rate at 71.68\%, followed by \texttt{GPT} at 68.14\%
Among non-commercial models, \texttt{Qwen}, \texttt{Llama}, \texttt{Janus}, \texttt{VL2}, and \texttt{R1} are 64.83\%, 59.70\%, 58.98\%, 51.29\%, and 57.70\%, respectively. As the temperature increases, the general trend of models is declining, with R1 showing slight instability. The differences across temperature settings are 0.4\% (\texttt{Gemini}), 1.19\% (\texttt{Qwen}), 1.62\% (\texttt{Llama}), 4.51\% (\texttt{Janus}), 10.85\% (\texttt{VL2}), and 1.8\% (\texttt{R1}), respectively. \texttt{VL2} exhibits the largest drop from 55.00\% to 44.15\%. In this setting, \texttt{Gemini} achieves the best performance with HTML inputs at 71.88\%. The results demonstrate that there is no big difference between open-source text-based LLMs and open-source vision-based LLMs. 
\looseness=-1

\PP{Screenshots vs. Logos vs. HTML}
LLMs perform worse with HTML than with screenshot or logo inputs, indicating that HTML alone lacks sufficient context for brand recognition. However, HTML still offers more cues than URLs, as reflected in improved performance.

\observ{Among individual inputs, screenshot inputs yield the best performance, and commercial models outperform non-commercial ones. URL inputs often contain irrelevant information, making brand detection difficult for LLMs. Furthermore, higher temperatures generally lead to worse performance. Therefore, setting the temperature to 0.0 is recommended, consistent with its role in controlling output randomness}

\subsection{Combination Inputs} \label{subsec:combination}
\begin{table}[!t]
    \centering
    \caption{\textbf{Brand Identification Performance Across Combined Inputs and Temperature Settings.} Bold values denote the highest identification rate within each input category.}
    \label{tab:result_combination}
    \resizebox{0.97\linewidth}{!}{
    \begin{tabular}{p{0.09\linewidth} p{0.08\linewidth} rrrrr r}
        \toprule
        \multirow{2}{*}{\textbf{Input}} & \multirow{2}{*}{\textbf{Model}} & \multicolumn{5}{c}{\textbf{Temperature}} & \multirow{2}{*}{\textbf{Avg.}}\\
        \cmidrule(lr){3-7}
        & & 0.0/0.1 & 0.3 & 0.5 & 0.7 & 1.0 \\\midrule
        \multirow{6}{*}{\makecell[l]{\textbf{Scr.}\\\textbf{URL}}} & GPT &  93.76 & NA & NA & NA & NA & 93.76\\
        & Gemini &  \textbf{94.59} & 94.54 & 94.54 & 94.47 & 94.51 & 94.53\\
        & Qwen &  87.05 & 86.92 & 86.84 & 86.46 & 85.45 & 86.54\\
        & Llama &  63.46 & 64.28 & 56.79 & 57.31 & 56.57 & 59.68\\
        & Janus &73.18 & 73.73 & 73.97 & 74.14 & 69.13 & 72.83\\
        & VL2 &  83.98  & 82.48  & 81.75  & 79.79  & 72.30  & 80.06\\\midrule
        \multirow{6}{*}{\makecell[l]{\textbf{Logo}\\\textbf{URL}}} & GPT &  86.57 & NA & NA & NA & NA & 86.57\\
        & Gemini &  \textbf{87.42} & 87.36 & 87.30 & 87.34 & 87.19 & 87.32\\
        & Qwen &  80.74 & 80.66 & 80.58 & 80.49 & 79.89 & 80.47\\
        & Llama &  74.10 & 70.39 & 72.94 & 71.71 & 70.06 & 71.84\\
        & Janus & 81.14 & 81.06 & 80.69 & 80.37 & 78.40 & 80.33\\
        & VL2 &  80.34 & 80.18 & 79.79 & 79.01 & 76.03 & 79.07\\\midrule
        \multirow{6}{*}{\raggedright\makecell[l]{\textbf{Scr.}\\\textbf{URL}\\\textbf{HTML}}} & GPT &  91.84 & NA & NA & NA & NA& 91.84\\
        & Gemini &  \textbf{93.72} & NA & NA & NA & NA & 93.72\\
        & Qwen & 84.98 & NA & NA & NA & NA & 84.98\\
        & Llama  & 74.16 & NA & NA & NA & NA & 74.16\\
        & Janus  & 49.32 & NA & NA & NA & NA & 49.32\\
        & VL2  & 76.81 & NA & NA & NA & NA & 76.81\\\bottomrule
    \end{tabular}
    }
\end{table}

We further examine the impact of \textit{multimodal} input combinations on phishing brand identification, including URLs with screenshots, URLs with logos, and URLs with both HTML content and screenshots, as summarized in~\autoref{tab:result_combination}.

\subsubsection{Results of URL-Screenshot}
With URL-screenshot inputs, \texttt{Gemini} achieves the best performance, with an average brand identification rate of 94.53\% across temperature settings, followed by \texttt{GPT} at 93.76\%. \texttt{Qwen} and \texttt{VL2} attain average scores of 86.54\% and 80.06\%, while \texttt{Janus} and \texttt{Llama} are 72.83\% and 59.68\%, respectively. As temperature increases, \texttt{Gemini} and \texttt{Qwen} remain relatively stable, showing 0.12\% and 1.6\% variation, respectively. In contrast, \texttt{Llama}, \texttt{Janus}, and \texttt{VL2} are sensitive, with fluctuations of 7.71\%, 5.01\%, and 11.68\%, respectively.

\PP{Screenshot vs. URL-Screenshot}Comparing results between screenshot alone and URL-screenshot inputs, \texttt{Gemini}'s average performance slightly improves from 94.41\% to 94.53\% (0.12\%{\small\faLongArrowUp}) when combining URLs as inputs, whereas \texttt{GPT} experiences a minor decrease from 93.95\% to 93.76\% (0.19\%{\small\faLongArrowDown}). \texttt{Qwen} decreases significantly from 91.01\% to 86.54\% (4.47\%{\small\faLongArrowDown}), \texttt{Llama} from 87.67\% to 59.68\% (27.99\%{\small\faLongArrowDown}), \texttt{Janus} from 77.86\% to 72.83\% (5.03\%{\small\faLongArrowDown}), and \texttt{VL2} from 84.05\% to 80.06\% (3.99\%{\small\faLongArrowDown}).
These results suggest that incorporating additional URLs has little impact on \texttt{GPT} and \texttt{Gemini}, slightly reduces performance for \texttt{Qwen}, \texttt{Janus}, and \texttt{VL2}, but significantly degrades performance for \texttt{Llama}, indicating that \texttt{Llama} is more sensitive to additional URL inputs.

\PP{URL vs. URL-Screenshot}When comparing URL-only inputs with URL-screenshot inputs, all LLMs significantly benefit from the inclusion of screenshot information, underscoring the critical role of screenshots in brand identification.

\observ{Commercial LLMs remain stable across temperature settings with URL-screenshot inputs, while \texttt{Llama}, \texttt{Janus}, and \texttt{VL2} exhibit high sensitivity. Additional URLs have little effect on commercial LLMs, cause a slight performance drop for \texttt{Janus} and \texttt{VL2}, and lead to significant degradation for \texttt{Qwen} and \texttt{Llama}. Furthermore, LLMs benefit from the inclusion of screenshots compared to URL-only inputs, reaffirming the crucial role of screenshots}

\subsubsection{Result of URL-Logo}
With URLs and logos as input, \texttt{Gemini} delivers the highest average performance at 87.32\% and an excellent stability with only 0.23\% variations at different temperatures. \texttt{GPT} follows with 86.57\%, while \texttt{Qwen}, \texttt{Janus}, \texttt{VL2}, and \texttt{Llama} achieve 80.47\%, 80.33\%, 79.07\%, and 71.84\%, with corresponding variations of 0.85\%, 2.74\%, 4.31\%, and 4.04\%, respectively.

\PP{Logo vs. URL-Logo}Compared to logo inputs, \texttt{GPT} (3.85\%{\small\faLongArrowUp}), \texttt{Gemini} (3.59\%{\small\faLongArrowUp}), \texttt{Qwen} (0.33\%{\small\faLongArrowUp}), \texttt{Janus} (1.82\%{\small\faLongArrowUp}), and \texttt{VL2} (3.17\%{\small\faLongArrowUp}) increase when combined with additional URLs. It indicates that URLs provide extra information beyond logos for detecting brands in most models, though using URL-only inputs remains weak. However, when combining the results for screenshot-only vs. URL-screenshot, we find that screenshots already convey richer information than logos, and adding URLs may introduce noise and decrease the results.

\PP{URL-Screenshot vs. URL-Logo}Compared to URL-screenshot, \texttt{GPT} drops from 93.76\% to 86.57\% (7.19\%{\small\faLongArrowDown}), \texttt{Gemini} from 94.53\% to 87.32\% (7.21\%{\small\faLongArrowDown}), \texttt{Qwen} from 86.54\% to 80.47\% (6.07\%{\small\faLongArrowDown}), and \texttt{VL2} from 80.06\% to 79.07\% (0.99\%{\small\faLongArrowDown}) when using URL-logo inputs. 
It shows that screenshots carry more convincing brand-related information than logos for these LLMs.
On the contrary, \texttt{Janus} improves by 7.50\%. Combined with results between screenshot-only vs. logo-only inputs, it shows that \texttt{Janus} is more suitable for logo-based inputs than screenshot-based inputs, particularly URL-logo.

\texttt{Llama} deviates from the previous trends, declining 7.21\% from logo to URL-logo but increasing 12.16\% from URL-screenshot to URL-logo, indicating the sensitivity to URLs. Furthermore, URL-screenshot inputs are more easily triggering safety mechanisms than URL-logos, causing \texttt{Llama} to refuse to respond. Specifically, 49.71\% of failed samples are refused on URL-screenshot inputs, while 35.26\% of samples are refused on URL-logo inputs. This indicates that URL-screenshot inputs make \texttt{Llama} behave more conservatively.

\observ{URLs provide complementary brand information beyond what logos offer. However, URLs may introduce noise when combined with screenshots that already convey rich and convincing brand information. \texttt{Llama} is conservative with URL-screenshot, while \texttt{Janus} appears better suited to URL-logo inputs}

\subsubsection{Result of URL-Screenshot-HTML}\texttt{Gemini} demonstrates the highest performance at 93.72\% when taking URLs, screenshots, and HTML as input with a temperature setting of 0.0, followed by \texttt{GPT} at 91.84\%, \texttt{Qwen} at 84.98\%, \texttt{VL2} at 76.81\%, \texttt{Llama} at 74.16\%, and \texttt{Janus} at 49.32\%.
Compared to URL-screenshot inputs, adding HTML reduces average performance for \texttt{GPT} (1.92\%{\small\faLongArrowDown}), \texttt{Gemini} (0.81\%{\small\faLongArrowDown}), \texttt{Qwen} (1.56\%{\small\faLongArrowDown}), \texttt{Janus} (23.51\%{\small\faLongArrowDown}), and \texttt{VL2} (3.25\%{\small\faLongArrowDown}). \texttt{Llama} is the only model that benefits (14.48\%), suggesting that HTML offers complementary brand information for \texttt{Llama} but may introduce conflicting signals with screenshots for other models. 
Conversely, compared to HTML-only inputs, \texttt{GPT} (23.27\%{\small\faLongArrowUp}), \texttt{Gemini} (21.84\%{\small\faLongArrowUp}), \texttt{Qwen} (19.71\%{\small\faLongArrowUp}), \texttt{Llama} (13.83\%{\small\faLongArrowUp}), and \texttt{VL2} (21.81\%{\small\faLongArrowUp}) all benefit from the inclusion of URLs and screenshots, whereas \texttt{Janus} (11.26\%{\small\faLongArrowDown}) shows a decrease.
Finally, compared to screenshot-only inputs, all models experience performance degradation, indicating that screenshots are the most informative component, while URLs and HTML tend to introduce noisy information. As shown in~\autoref{subsec:depth_single}, textual content from URLs and HTML can conflict with screenshots, leading to mispredictions.

\observ{Screenshots are the most informative component, while URLs and HTML tend to introduce noisy information}

\subsection{One-shot Prompting Technique} \label{subsec:oneshot}
\begin{table}[!t]
    \centering
    \caption{\textbf{Brand Identification Performance Across One-shot Inputs and Temperature Settings.} Bold values denote the highest identification rate within each input category.}
    \label{tab:result_oneshot}
    \resizebox{\linewidth}{!}{
    \begin{tabular}{p{0.13\linewidth} l rrrrr l}
        \toprule
        \multirow{2}{*}{\textbf{Input}} & \multirow{2}{*}{\textbf{Model}} & \multicolumn{5}{c}{\textbf{Temperature}} & \multirow{2}{*}{\textbf{Avg.}}\\
        \cmidrule(lr){3-7}
        & & 0.0/0.1 & 0.3 & 0.5 & 0.7 & 1.0 \\\midrule
        \multirow{6}{*}{\makecell[l]{\textbf{Scr.}\\\textbf{One-shot}}} & GPT &  \textbf{93.39} & NA & NA & NA & NA & 93.39\\
        & Gemini & 92.63 & 92.67 & 92.55 & 92.56 & 92.47 & 92.58\\
        & Qwen & 90.13 & 90.24 & 90.23 & 89.93 & 89.59 & 90.02\\
        & Janus &  80.17 & 80.24 & 79.84 & 79.57 & 77.12 & 79.39\\
        & VL2 &  70.11 & 70.50 & 73.10 & 68.26 & 57.75 & 67.94\\\midrule
        \multirow{6}{*}{\makecell[l]{\textbf{Logo}\\\textbf{One-shot}}} & GPT & 81.06 & NA & NA & NA & NA & 81.06\\
        & Gemini &  82.75 & 82.79 & 82.79 & 82.78 & \textbf{82.83} & 82.79\\
        & Qwen &  80.26 & 80.19 & 80.16 & 79.98 & 79.82 & 80.08\\
        & Janus & 78.09 & 77.78 & 76.35 & 72.13 & 65.65 & 74.00\\
        & VL2 &  76.24 & 76.32 & 76.05 & 75.81 & 71.51 & 75.19\\\midrule
        \multirow{6}{*}{\makecell[l]{\textbf{URL}\\\textbf{One-shot}}} & GPT &  \textbf{36.32} & NA & NA & NA & NA & 36.32\\
        & Gemini &  29.37 & 29.45 & 29.46 & 29.35 & 29.25 & 29.38\\
        & Qwen &  25.33 & 25.16 & 24.82 & 24.04 & 22.93 & 24.46\\
        & Llama &  32.03 & 31.89 & 31.72 & 30.60 & 30.07 & 31.26\\
        & Janus &  20.63 & 20.37 & 19.45 & 18.70 & 16.05 & 19.04\\
        & VL2 &  22.00 & 21.64 & 21.02 & 20.34 & 18.66 & 20.73\\
        & R1 & 15.76 & 15.90 & 16.00 & 15.74 & 15.20 & 15.72 \\\bottomrule        
    \end{tabular}
    }
\end{table}

\PP{Screenshot-One-shot Results.}With a screenshot example as guidance, \texttt{GPT} shows the best performance of 93.39\%, followed by \texttt{Gemini} with an average value of 92.58\%, \texttt{Qwen} at 90.02\%, \texttt{Janus} at 79.39\%, and \texttt{VL2} at 67.94\% in \autoref{tab:result_oneshot}. Compared to the screenshot-only input, the performance of \texttt{GPT}, \texttt{Gemini}, \texttt{Qwen}, and \texttt{VL2} decreases by 0.56\%, 1.83\%, 0.09\%, and 16.11\%, respectively, when switching the input from screenshot-only to screenshot-one-shot. In contrast, \texttt{Janus} improves by 1.53\%. It suggests that the guided example helps \texttt{Janus} make better predictions, but introduces noise and reduces the accuracy for others.

\PP{Logo-One-shot Results.}With a logo example as guidance, \texttt{Gemini} indicates the best average performance across temperatures at 82.79\%, followed by \texttt{GPT} (81.06\%) and \texttt{Qwen} (80.08\%). \texttt{Janus} and \texttt{VL2} trail behind with averages of 74.00\% and 75.19\%, respectively. Compared to screenshot-one-shot results, all LLMs decrease while \texttt{VL2} improves from 67.94\% to 75.19\%. It indicates that an example can guide most LLMs with screenshot inputs better than logos. 
Compared to logo results, the performance of \texttt{GPT} (1.66\%{\small\faLongArrowDown}), \texttt{Gemini} (0.94\%{\small\faLongArrowDown}), \texttt{Qwen} (0.06\%{\small\faLongArrowDown}), \texttt{Janus} (4.51\%{\small\faLongArrowDown}), and \texttt{VL2} (0.71\%{\small\faLongArrowDown}) decreases. It shows that logo guidance has little impact on brand prediction for these LLMs in this dataset.

\PP{URL-One-shot Results.}Using a URL example as guidance yields weak performance, with all results between 15\% and 37\%. Compared to plain URL inputs, most LLMs benefit from example-based guidance. However, \texttt{Gemini}, \texttt{Janus}, and \texttt{R1} exhibit slight performance decreases. Regardless, both URL and URL-One-shot results remain far below other input types, showing that URL-based inputs alone are insufficient for reliable prediction.

\observ{URL inputs, whether standalone or guided by examples, consistently yield poor performance. One-shot guidance is more effective with screenshots than with logos. However, one-shot guidance generally has little effect or even degrades performance for most LLMs compared to screenshot or logo inputs alone}

\section{In-Depth Analysis of Brand Misclassification} 
\label{sec:analysis}

\begin{table*}[!t]
    \centering
    \caption{\textbf{Category Occurrence in Failed Samples Across Inputs.} Failure Total is the total number of failed samples. Valid Response refers to the number of failures containing at least one word from the six categories.
    Values under each category show the word count and its corresponding percentage out of the failed samples. Bold values indicate the highest percentage for LLMs.}
    \label{tab:failure_words}
    \resizebox{0.955\linewidth}{!}{
    \begin{tabular}{p{0.065\linewidth} p{0.06\linewidth} p{0.055\linewidth} r rrrrrr}
        \toprule
        \multirow{2}{*}{\textbf{Input}} & \multirow{2}{*}{\textbf{Model}} & \multirow{2}{*}{\makecell[c]{\textbf{Failure} \\\textbf{Total}}} & \multirow{2}{*}{\makecell[c]{\textbf{Valid$^*$}\\\textbf{Response}}} & \multicolumn{6}{c}{\textbf{Response within Word Category}}\\ \cmidrule(lr){5-10}
        & & & & \multicolumn{1}{c}{Auth.\&ID} & \multicolumn{1}{c}{Layout\&Pos.} & \multicolumn{1}{c}{Color\&Style} & \multicolumn{1}{c}{Inter. Element} & \multicolumn{1}{c}{Text\&Lang.} & \multicolumn{1}{c}{Media\&Graphics} \\\midrule
        \multirow{6}{*}{\textbf{Scr.}} 
        & GPT& 1,157 & 1,155 (99.83\%) & 788 (68.11\%) & 519 (44.86\%) & 757 (65.43\%) & 757 (65.43\%) & \textbf{1,025 (88.59\%)} & 635 (54.88\%)\\
        & Gemini & 1,061  & 1,026 (96.70\%) & 530 (49.95\%) & 238 (22.43\%) & 313 (29.50\%) & 484 (45.62\%) & \textbf{748 (70.50\%)} & 235 (22.15\%)\\
        & Qwen &1,536& 1,524 (99.22\%) & 972 (63.28\%) & 696 (45.31\%) & 852 (55.47\%) & 942 (61.33\%) & \textbf{1,307 (85.09\%)} & 579 (37.70\%) \\
        & Llama & 2,055& 1,988 (96.74\%)& 1,092 (53.14\%) & 739 (35.96\%) & 836 (40.68\%) & 712 (34.65\%) & \textbf{1,457 (70.90\%)} & 947 (46.08\%) \\
        & Janus & 3,834& 3,826 (99.79\%)& 2,870 (74.86\%) & 2,974 (77.57\%) & 2,986 (77.88\%) & 2,876 (75.01\%) & \textbf{3,196 (83.36\%)} & 1,901 (49.58\%) \\
        & VL2  & 2,332 & 1,086 (46.57\%)& 566 (24.27\%) & 427 (18.31\%) & 278 (11.92\%) & 481 (20.63\%) & \textbf{759 (32.55\%)} & 336 (14.41\%)\\\midrule
        \multirow{6}{*}{\textbf{Logo}}  
        & GPT  & 3,306& 2,864 (86.63\%)& 2,348 (71.02\%) & 1,339 (40.50\%) & \textbf{2,797 (84.60\%)} & 1,514 (45.80\%) & 1,925 (58.23\%) & 2,286 (69.15\%) \\
        & Gemini   & 3,108& 2,668 (85.84\%)& 1,877 (60.39\%) & 825 (26.54\%) & \textbf{2,083 (67.02\%)} & 933 (30.02\%) & 1,767 (56.85\%) & 1,645 (52.93\%) \\
        & Qwen  & 3,808& 3,371 (88.52\%)& 2,179 (57.22\%) & 1,671 (43.88\%) & 2,990 (78.52\%) & 1,573 (41.31\%) & \textbf{3,040 (79.83\%)} & 2,773 (72.82\%) \\
        & Llama & 3,961& 3,496 (88.26\%) & 2,435 (61.47\%) & 1,572 (39.69\%) & \textbf{2,784 (70.29\%)} & 1,915 (48.35\%) & 2,029 (51.22\%) & 1,756 (44.33\%) \\
        & Janus & 4,124& 3,684 (89.33\%) & 2,588 (62.75\%) & 2,371 (57.49\%) & \textbf{3,230 (78.32\%)} & 1,304 (31.62\%) & 2,370 (57.47\%) & 2,394 (58.05\%) \\
        & VL2 & 4,780& 4,193 (87.72\%) & 2,131 (44.58\%) & 2,189 (45.79\%) & \textbf{3,145 (65.79\%)} & 1,560 (32.64\%) & 2,519 (52.70\%) & 1,824 (38.16\%) \\\midrule
        \multirow{6}{*}{\textbf{URL}} 
        & GPT& 13,619& 13,565 (99.60\%) & \textbf{12,159 (89.28\%)} & 5,671 (41.64\%) & 5,271 (38.70\%) & 5,967 (43.81\%) & 7,080 (51.99\%) & 4,037 (29.64\%) \\
        & Gemini  & 13,356& 12,592 (92.89\%) & \textbf{9,728 (72.84\%)} & 2,216 (16.59\%) & 1,787 (13.38\%) & 2,300 (17.22\%) & 5,421 (40.59\%) & 3,748 (28.06\%)\\
        & Qwen & 16,453& 16,443 (99.94\%) & 6,251 (37.99\%) & 6,414 (38.98\%) & 7,715 (46.89\%) & 8,741 (53.13\%) & \textbf{15,666 (95.22\%)} & 7,201 (43.77\%) \\
        & Llama & 14,179 & 14,132 (99.67\%) & \textbf{12,481 (88.02\%)} & 3,513 (24.78\%) & 3,341 (23.56\%) & 4,731 (33.37\%) & 11,160 (78.71\%) & 6,407 (45.19\%)\\
        & Janus & 14,467 & 14,184 (98.04\%) & 8,488 (58.67\%) & 2,011 (13.90\%) & 769 \phantom{0}(5.32\%) & 2,893 (20.00\%) & \textbf{12,076 (83.47\%)} & 6,953 (48.06\%) \\
        & VL2 & 15,243& 12,969 (85.08\%)& 4,432 (29.08\%) & 3,210 (21.06\%) & 844 \phantom{0}(5.54\%) & 2,891 (18.97\%) & \textbf{7,905 (51.86\%)} & 7,434 (48.77\%) \\ 
        & R1 & 15,917& 10,831 (68.05\%)  & \textbf{6,092 (38.27\%)} & 4,130 (25.95\%) & 757 \phantom{0}(4.76\%) & 2,283 (14.34\%) & 5,579 (35.05\%) & 2,382 (14.97\%)  \\\midrule
        \multirow{6}{*}{\textbf{HTML}} 
        & GPT & 6,095& 6,088 (99.89\%)  & 2,853 (46.81\%) & 2,083 (34.18\%) & 1,170 (19.20\%) & 3,786 (62.12\%) & \textbf{5,755 (94.42\%)} & 2,039 (33.45\%) \\
        & Gemini & 5,380 & 5,356 (99.55\%) & 1,999 (37.16\%) & 1,681 (31.25\%) & 1,542 (28.66\%) & 2,355 (43.77\%) & \textbf{4,831 (89.80\%)} & 2,171 (40.35\%) \\
        & Qwen  & 6,645& 6,642 (99.95\%)  & 4,498 (67.69\%) & 3,429 (51.60\%) & 1,637 (24.64\%) & 4,035 (60.72\%) & \textbf{6,235 (93.83\%)} & 3,670 (55.23\%)\\
        & Llama  & 7,589& 6,920 (91.18\%) & 3,414 (44.99\%) & 3,718 (48.99\%) & 2,283 (30.08\%) & 3,942 (51.94\%) & \textbf{6,137 (80.87\%)} & 4,022 (53.00\%) \\
        & Janus & 7,541 & 7,235 (95.94\%)& 2,168 (28.75\%) & 1,348 (17.88\%) & 1,079 (14.31\%) & 2,201 (29.19\%) & \textbf{5,698 (75.56\%)} & 2,646 (35.09\%)  \\
        & VL2 & 8,609& 7,211 (83.76\%) & 1,751 (20.34\%) & 1,432 (16.63\%) & 659 \phantom{0}(7.65\%) & 3,290 (38.22\%) & \textbf{4,197 (48.75\%)} & 3,785 (43.97\%)  \\
        & R1  & 8,062& 6,569 (81.48\%)& 3,042 (37.73\%) & 2,316 (28.73\%) & 719 \phantom{0}(8.92\%) & 3,490 (43.29\%) & \textbf{5,882 (72.96\%)} & 2,555 (31.69\%) \\\midrule
        \multirow{6}{*}{\makecell[l]{\textbf{Scr.}\\\textbf{URL}}} 
        & GPT & 1,193& 1,192 (99.92\%) & 1,038 (87.01\%) & 704 (59.01\%) & 829 (69.49\%) & 863 (72.34\%) & \textbf{1,067 (89.44\%)} & 606 (50.80\%) \\
        & Gemini & 1,035& 977 (94.40\%) & 591 (57.10\%) & 210 (20.29\%) & 245 (23.67\%) & 466 (45.02\%) & \textbf{659 (63.67\%)} & 302 (29.18\%)  \\
        & Qwen & 2,477& 2,474 (99.88\%) & 1,746 (70.49\%) & 1,147 (46.31\%) & 1,108 (44.73\%) & 1,545 (62.37\%) & \textbf{2,206 (89.06\%)} & 1,431 (57.77\%)  \\
        & Llama & 6,990& 6,032 (86.29\%) & 2,809 (40.19\%) & 1,753 (25.08\%) & 1,696 (24.26\%) & 3,108 (44.46\%) & \textbf{4,134 (59.14\%)} & 2,629 (37.61\%) \\
        & Janus  & 5,130 & 3,746 (73.02\%) & 1,244 (24.25\%) & 572 (11.15\%) & 678 (13.22\%) & 899 (17.52\%) & \textbf{2,866 (55.87\%)} & 893 (17.41\%)  \\
        & VL2 & 3,064 & 990 (32.31\%) & 322 (10.51\%) & 220 \phantom{0}(7.18\%) & 202 \phantom{0}(6.59\%) & 324 (10.57\%) & \textbf{627 (20.46\%)} & 391 (12.76\%)  \\ \midrule
        \multirow{6}{*}{\makecell[l]{\textbf{Logo} \\\textbf{URL}}} 
        & GPT  & 2,570 & 2,134 (83.04\%) & \textbf{2,012 (78.29\%)} & 1,232 (47.94\%) & 1,927 (74.98\%) & 1,682 (65.45\%) & 1,892 (73.62\%) & 1,889 (73.50\%) \\
        & Gemini  & 2,407 & 1,921 (79.81\%) & \textbf{1,341 (55.71\%)} & 473 (19.65\%) & 833 (34.61\%) & 756 (31.41\%) & 1,362 (56.58\%) & 859 (35.69\%)\\
        & Qwen & 3,685 & 3,247 (88.11\%) & 2,562 (69.53\%) & 1,969 (53.43\%) & 2,300 (62.42\%) & 2,625 (71.23\%) & 2,610 (70.83\%) & \textbf{2,660 (72.18\%)} \\
        & Llama  & 4,955 & 4,310 (86.98\%)& 3,147 (63.51\%) & 1,906 (38.47\%) & 2,350 (47.43\%) & 2,668 (53.84\%) & \textbf{3,287 (66.34\%)} & 2,340 (47.23\%)  \\
        & Janus  & 3,609& 3,123 (86.53\%) & 2,040 (56.53\%) & 2,112 (58.52\%) & 1,955 (54.17\%) & 1,011 (28.01\%) & \textbf{2,204 (61.07\%)} & 1,621 (44.92\%) \\
        & VL2 & 3,762& 3,029 (80.52\%)& 2,076 (55.18\%) & 1,886 (50.13\%) & 1,996 (53.06\%) & 1,329 (35.33\%) & 1,997 (53.08\%) & \textbf{2,213 (58.83\%)} \\ \midrule
        \multirow{5}{*}{\makecell[l]{\textbf{Scr.}\\\textbf{URL}\\\textbf{HTML}}} 
        & GPT  & 1,561& 1,560 (99.94\%) & 1,414 (90.58\%) & 1,042 (66.75\%) & 934 (59.83\%) & 1,204 (77.13\%) & \textbf{1,550 (99.30\%)} & 898 (57.53\%) \\
        & Gemini & 1,202 & 1,193 (99.25\%) & 688 (57.24\%) & 411 (34.19\%) & 231 (19.22\%) & 582 (48.42\%) & \textbf{1,120 (93.18\%)} & 479 (39.85\%) \\
        & Qwen  & 2,874 & 2,873 (99.97\%)& 2,188 (76.13\%) & 1,806 (62.84\%) & 1,323 (46.03\%) & 2,244 (78.08\%) & \textbf{2,849 (99.13\%)} & 2,019 (70.25\%) \\
        & Llama & 4,944 & 4,924 (99.60\%)& 2,731 (55.24\%) & 2,524 (51.05\%) & 2,202 (44.54\%) & 3,175 (64.22\%) & \textbf{4,663 (94.32\%)} & 3,984 (80.58\%)  \\
        & Janus  & 9,696 & 9,588 (98.89\%)& 3,308 (34.12\%) & 2,099 (21.65\%) & 2,138 (22.05\%) & 3,306 (34.10\%) & \textbf{9,081 (93.66\%)} & 7,395 (76.27\%)\\
        & VL2 & 4,437  & 2,264 (51.03\%)  & 1,009 (22.74\%) & 554 (12.49\%) & 275 \phantom{0}(6.20\%) & 754 (16.99\%) & \textbf{1,576 (35.52\%)} & 876 (19.74\%) \\ \midrule
        \multirow{5}{*}{\makecell[l]{\textbf{Scr.}\\\textbf{One-shot}}}
        & GPT& 1,265  & 1,264 (99.92\%) & 820 (64.82\%) & 460 (36.36\%) & 808 (63.87\%) & 845 (66.80\%) & \textbf{1,052 (83.16\%)} & 504 (39.84\%) \\
        & Gemini & 1,409 & 1,394 (98.94\%) & 910 (64.58\%) & 282 (20.01\%) & 377 (26.76\%) & 852 (60.47\%) & \textbf{1,092 (77.50\%)} & 327 (23.21\%) \\
        & Qwen & 1,889 & 1,888 (99.95\%)& 1,355 (71.73\%) & 885 (46.85\%) & 1,205 (63.79\%) & 1,524 (80.68\%) & \textbf{1,629 (86.24\%)} & 817 (43.25\%) \\
        & Janus & 3,794  & 3,721 (98.08\%) & 2,242 (59.09\%) & 1,607 (42.36\%) & \textbf{2,613 (68.87\%)} & 2,556 (67.37\%) & 2,047 (53.95\%) & 1,475 (38.88\%)\\
        & VL2  & 5,718  & 5,666 (99.09\%)& \textbf{5,617 (98.23\%)} & 569 \phantom{0}(9.95\%) & 2,075 (36.29\%) & 5,317 (92.99\%) & 199 \phantom{0}(3.48\%) & 102 (1.78\%) \\ \midrule
        \multirow{5}{*}{\makecell[l]{\textbf{Logo}\\\textbf{One-shot}}} 
        & GPT   & 3,624 & 3,188 (87.97\%) & 2,281 (62.94\%) & 1,535 (42.36\%) & \textbf{3,106 (85.71\%)} & 1,453 (40.09\%)         & 2,191 (60.46\%) & 2,593 (71.55\%) \\
        & Gemini& 3,300 & 2,847 (86.27\%) & 1,746 (52.91\%) & 999 (30.27\%)   & \textbf{2,379 (72.09\%)} & 905 (27.42\%)           & 2,123 (64.33\%) & 1,393 (42.21\%) \\
        & Qwen  & 3,777 & 3,341 (88.46\%) & 2,256 (59.73\%) & 2,253 (59.65\%) & \textbf{3,210 (84.99\%)} & 2,130 (56.39\%)         & 2,704 (71.59\%) & 2,769 (73.31\%) \\
        & Janus & 4,192 & 3,749 (89.43\%) & 917 (21.88\%)   & 3,193 (76.17\%) & 3,600 (85.88\%)          & 504 (12.02\%)           & \textbf{3,608 (86.07\%)} & 1,313 (31.32\%)\\
        & VL2   & 4,546 & 4,090 (89.97\%) & 502 (11.04\%)   & 3,844 (84.56\%) & \textbf{3,862 (84.95\%)} & 161 \phantom{0}(3.54\%) & 1,100 (24.20\%) & 3,656 (80.42\%)\\ \midrule
        \multirow{7}{*}{\makecell[l]{\textbf{URL}\\\textbf{One-shot}}} 
        & GPT       & 12,182& 12,154 (99.77\%)  & \textbf{10,750 (88.24\%)} & 4,434 (36.40\%) & 5,902 (48.45\%) & 5,526 (45.36\%) & 8,917 (73.20\%) & 3,915 (32.14\%)        \\
        & Gemini    & 13,512& 13,085 (96.84\%)  & \textbf{10,186 (75.38\%)} & 3,847 (28.47\%) & 2,391 (17.70\%) & 3,439 (25.45\%) & 7,126 (52.74\%) & 4,836 (35.79\%)        \\
        & Qwen      & 14,286& 13,979 (97.85\%)  & 5,520 (38.64\%) & 6,091 (42.64\%) & 10,522 (73.65\%) & 4,022 (28.15\%) & \textbf{11,848 (82.93\%)} & 5,182 (36.27\%)       \\
        & Llama     & 13,003& 12,950 (99.59\%)  & \textbf{10,848 (83.43\%)} & 4,693 (36.09\%) & 3,682 (28.32\%) & 4,408 (33.90\%) & 9,315 (71.64\%) & 5,241 (40.31\%)        \\
        & Janus     & 15,185& 15,051 (99.12\%)  & 13,161 (86.67\%) & 2,114 (13.92\%) & 9,292 (61.19\%) & 2,272 (14.96\%) & \textbf{14,013 (92.28\%)} & 3,511 (23.12\%)       \\
        & VL2       & 14,922& 12,515 (83.87\%)  & 5,004 (33.53\%) & 3,603 (24.15\%) & 805 \phantom{0}(5.39\%) & 2,150 (14.41\%) & \textbf{7,637 (51.18\%)} & 6,869 (46.03\%) \\ 
        & R1& 16,115& 7,510 (46.60\%)   & \textbf{5,113 (31.73\%)} & 3,270 (20.29\%) & 826 \phantom{0}(5.13\%) & 1,628 (10.10\%) & 3,697 (22.94\%) & 1,738 (10.78\%) \\ \bottomrule
        \multicolumn{10}{l}{$^*$Valid Response: the number of failures containing at least one word from the six categories.}
    \end{tabular}
    }
\end{table*}

Based on \autoref{tab:explanation_category}, we first analyze the word distribution of 19,131 outputs to identify general LLM trends. The statistics are summarized in~\autoref{tab:word_explanation} of~\aref{sec:total_word_distribution}. We then examine valid responses among failed examples. A response is considered valid if it contains at least one word from these six groups. Otherwise, we treat the response as irrelevant. We then calculate their word distributions in \autoref{tab:failure_words} and analyze potential reasons for failure associated with each input prompt.
We focus on results with $do\_sample$ set to $False$ (temperature of 0.1 for \texttt{Qwen}, 0.0 for others), where all models achieve the best or comparable performance, simplifying the analysis.

\subsection{Analysis of Single-Component Results} \label{subsec:depth_single}

\PP{Screenshot Results}Among screenshot inputs, \texttt{GPT}, \texttt{Gemini}, \texttt{Qwen}, \texttt{Llama}, and \texttt{Janus} produce high proportions of valid responses in failed samples ($\geq$96\%), showing that their explanations contain semantically relevant content. 
In contrast, \texttt{VL2} shows the lowest portion (46.57\%), which means that its failed outputs lack meaningful content. Moreover, most failures across models fall into the \textit{Text \& Language} category, suggesting a strong reliance on textual cues. 
For instance, as shown in~\autoref{fig:eg_scr_gpt_text}, \texttt{GPT} misidentifies \textit{Facebook} as \textit{CBS News} by focusing on textual content rather than visual branding.
The consistently low word-hit rates across all categories for \texttt{VL2} further confirm that its failures involve generating uninformative outputs rather than incorrect classifications.

\PP{Logo Results}All LLMs exhibit high validity between 85\% and 90\%. The \textit{Color \& Style} dominates across all models, particularly for \texttt{GPT} (84.60\%), and this trend is consistent with overall trends discussed in~\aref{subsec:depth_single_total}. In contrast, \textit{Layout \& Position} and \textit{Interactive Element} appear infrequently, which indicates that logos alone are insufficient for reliable brand inference. \texttt{VL2} shows low word-hit rates, which further indicates that its outputs often lack meaningful information.
\autoref{fig:eg_logo_gpt_color} shows a \texttt{GPT} failure, where it misidentifies \textit{Facebook} as \textit{Microsoft} based on visual theme and color.

\PP{URL Results}\texttt{GPT}, \texttt{Gemini}, \texttt{Qwen}, \texttt{Llama}, and \texttt{Janus} exhibit high validity rates ranging from 92\% to 100\%, while \texttt{VL2} and \texttt{R1} are relatively lower, indicating that these two models often refuse to respond or produce uninformative replies.
Consistent with overall URL trends discussed in~\aref{subsec:depth_single_total}, most failures are associated with \textit{Authentication \& Identity} and \textit{Text \& Language} categories. Specifically, \texttt{GPT} (89.28\%), \texttt{Gemini} (72.84\%), \texttt{Llama} (88.02\%), and \texttt{R1} (38.27\%) exhibit higher word coverage among \textit{Authentication \& Identity}, while \texttt{Qwen} (95.22\%), \texttt{Janus} (83.47\%), and \texttt{VL2} (51.86\%) are dominant in \textit{Text \& Language}. Other categories appear at comparatively lower rates. Despite high word coverage in these two categories, models still fail, suggesting that URLs contain many irrelevant or misleading words that disturb predictions. 
For instance, \texttt{GPT} fails on the URL (\url{deliveryagents.github.io/keghdyg}), targeting \textit{Facebook} but hosted on \textit{GitHub}. In this scenario, \texttt{GPT} is conservative and concludes with no clear brand.

\PP{HTML Results}\texttt{GPT}, \texttt{Gemini}, \texttt{Qwen}, \texttt{Llama}, and \texttt{Janus} exhibit high validity rates (91\%-100\%), while \texttt{VL2} and \texttt{R1} are slightly lower (81\%-84\%). HTML inputs yield preference for \textit{Text \& Language} category, with coverage rates of \texttt{GPT} (94.42\%), \texttt{Gemini} (89.80\%), \texttt{Qwen} (93.83\%), \texttt{Llama} (80.87\%), \texttt{Janus} (75.56\%), \texttt{VL2} (48.75\%), \texttt{R1} (72.96\%), respectively. Other categories show lower coverage, indicating that textual information in HTML among these failures often misleads LLMs.
For example, \texttt{GPT} misidentifies \textit{DHL} as \textit{Joinnow.Live} with HTML title contains \textit{Joinnow.Live Webinars}.


\observ{\texttt{GPT}, \texttt{Gemini}, \texttt{Qwen}, \texttt{Llama}, and \texttt{Janus} consistently produce semantically relevant outputs, while \texttt{VL2} and \texttt{R1} often fail with uninformative responses.
Failures with screenshot inputs are closely related to \textit{Text \& Language}, logo inputs emphasize challenges in \textit{Color \& Style}, and URL and HTML inputs often contain misleading words that cause mispredictions}

\subsection{Analysis of Combined-Component Results} \label{subsec:depth_combination}

\PP{Screenshot-URL Results}\texttt{GPT}, \texttt{Gemini}, and \texttt{Qwen} exhibit high validity rates ranging from 94.40\% to 99.92\%, \texttt{Llama} and \texttt{Janus} exhibit slightly lower coverage rates of 86.29\% and 73.02\%, respectively, while \texttt{VL2} shows the lowest rate of 32.31\%, indicating that its outputs contain lots of useless information, consistent with observations obtained from screenshot failure analysis.
Screenshot-URL inputs show higher coverage of \textit{Text \& Language} across all models, similar to screenshot input failures. Adding URLs makes \texttt{Gemini} (6.83\%{\small\faLongArrowDown}), \texttt{Llama} (11.76\%{\small\faLongArrowDown}), \texttt{Janus} (27.49\%{\small\faLongArrowDown}), and \texttt{VL2} (12.09\%{\small\faLongArrowDown}) decrease coverage, while \texttt{GPT} (0.85\%{\small\faLongArrowUp}) and \texttt{Qwen} (3.97\%{\small\faLongArrowUp}) slightly increase coverage for \textit{Text \& Language}. Overall, the dominant trend is a decline, suggesting that the addition of URLs pushes LLMs to focus more on textual content.

\PP{Logo-URL Results}All models show valid response rates around 79\%-87\%, slightly lower than logo inputs, which range from 85\%-90\%. Although logo inputs primarily trigger failures in the \textit{Text \& Language} category, logo-URL inputs introduce more diversity. Specifically, \texttt{GPT} (78.29\%) and \texttt{Gemini} (55.71\%) maintain strong trends toward failures in \textit{Authentication \& Identity}, \texttt{Qwen} (72.18\%) and \texttt{VL2} exhibit more in \textit{Media \& Graphics}, while \texttt{Llama} and \texttt{Janus} primarily fail in \textit{Text \& Language}. These suggest that combining URLs with logos introduces new sources of ambiguity that affect model reasoning across multiple semantic dimensions.

\PP{Screenshot-URL-HTML Results}Combining screenshots with URLs and HTML leads all models to fail mainly in the \textit{Text \& Language} category. This proportion increases across all LLMs compared to screenshot or screenshot-URL inputs, and also rises compared to HTML except for \texttt{VL2}, which decreases from 48.75\% to 35.52\%. However, a high failure concentration in this category does not guarantee success, implying the textual noise.

\observ{Although multimodal inputs have the potential to enrich LLM predictions, they also introduce noisy or redundant information that can reduce model performance. High coverage of textual content does not guarantee successful identification}

\subsection{Analysis of One-shot Results} \label{subsec:depth_oneshot}

\autoref{tab:failure_words} shows that one-shot guidance generally boosts valid response rates for screenshot-one-shot inputs, with the exception of \texttt{Janus} (1.71\%{\small\faLongArrowDown}), which exhibits a slight decrease compared to the screenshot-only input. \texttt{VL2} achieves the highest improvement, rising from 46.57\% to 99.09\%, primarily driven by \textit{Authentication \& Identity} category. However, a large portion of failed responses are useless, as reflected by the high occurrence but low detection accuracy.
Logo one-shots also increase or maintain comparable hit rates across models. For URL one-shot inputs, most models perform comparably, though \texttt{R1} drops from 68.05\% to 46.60\%. The overall failure patterns in one-shot settings are similar to those observed with their respective single inputs. 

\observ{One-shot guidance improves valid response rates across most input types and models, but high validity and coverage do not necessarily yield better identification accuracy. Failure patterns remain similar to single-input scenarios, underscoring persistent model limitations}

\section{Discussion} 
\label{sec:limitation}
While this study provides valuable insights into the effectiveness of LLM-based phishing detectors, we acknowledge several limitations that present opportunities for future research. 
First, our evaluation is based on the APWG eCX dataset and its corresponding benign brand dataset. Incorporating additional real-world datasets would better capture the diversity of phishing strategies across different platforms and regions. Future work could explore additional datasets such as OpenPhish~\cite{OpenPhis48:online} and PhishTank~\cite{PhishTan70:online} to improve generalizability and robustness across different threat landscapes. 
Second, LLM outputs are diverse and often fail to strictly follow the predefined rules under zero-shot and one-shot settings. This inconsistency poses challenges for automated evaluation and downstream integration. Therefore, future work can fine-tune LLMs to better restrict the outputs of models with task-specific formats. 
%
%

Based on our findings, we recommend using screenshots as the primary input, supplemented with URLs and HTML for additional context, and adopting lower temperature settings when designing phishing detection models. Carefully designed detection models that utilize brand information can achieve performance comparable to commercial LLMs with simple detection pipelines.

\section{Related Work} \label{sec:related_work}

\PP{Deep Learning-based Approaches}Deep learning-based approaches have evolved with the integration of visual information for phishing detection~\cite{Liu2022PhishIntention, Lin2021Phishpedia}. These methods detect brand intent and potential credential theft by comparing the input against a predefined list of trusted references, which include authentic domains and their associated visual or textual features. However, these works are constrained by their reliance on predefined reference lists.

\PP{LLMs-based Approaches}Recent works explore the application of LLMs in phishing detection. Koide \textit{et al.}~\cite{koide2024ChatSpamDetector} use LLMs to analyze email contents to detect phishing emails and give users explanations. 
Lee \textit{et al.}~\cite{lee2024mmlm} use screenshots and HTML as input to identify the brand of given websites for detecting phishing. Liu \textit{et al.}~\cite{Liu2024PhishLLM} leverage LLMs to infer brand based on both captions and accompanying text among logos. However, these works are insufficiently exploring which types of information are effective for LLM-based phishing detection, or how parameters influence performance.

\PP{Assessment of Phishing Detection Models}Recent studies~\cite{kulkarni2024mltollm,Liu2024PhishLLM,lee2024mmlm,Ji2025Usenix} have assessed phishing detection models from various aspects. Ji~\textit{et al.}~\cite{Ji2025Usenix} evaluated the effectiveness and robustness of popular visual similarity-based anti-phishing models using a 451k real-world phishing dataset. 
Aditya~\textit{et al.}~\cite{kulkarni2024mltollm} evaluated the robustness of traditional models and one multimodal LLM-based phishing detector using HTML and screenshots from generated adversarial phishing webpages.
Lee~\textit{et al.}~\cite{lee2024mmlm} evaluated the efficacy of Gemini, GPT, and Claude for detecting brands via HTML and screenshot inputs. 
%
However, these studies lack a comprehensive evaluation across diverse input modalities (\eg, screenshots, logos, HTML, URLs, and their combinations) along with parameters, and do not explore the underlying causes behind LLMs in phishing detection. 

\section{Conclusion} 
\label{sec:conclusion}
We evaluated seven popular LLMs on 19,131 real-world phishing and 243 benign websites to explore the effectiveness of LLMs compared to DL models and the impact of prompt engineering, \textit{individual} and \textit{multimodal} input components, and various temperature settings. 
We identified optimal configurations for achieving the best performance across different LLMs. 
Our analysis showed that LLMs achieved superior true positive rates, while DL models yielded higher true negative rates in phishing detection. For brand identification, lower temperatures generally yielded better results, particularly with screenshot inputs. 
Failures with screenshots or HTML inputs are often tied to textual content, whereas logo-related failures are primarily associated with color and style features. 
These observations are essential for developing more effective phishing detection systems with interpretable outputs.

\newpage
\section*{Ethical Considerations} \label{sec:ethics}

In this work, we focus on evaluating the performance of various LLMs and deep learning-based phishing detection models. In line with the principles of the Menlo Report, we carefully consider the potential ethical implications.

\PP{Stakeholders}Primary stakeholders of this research include: 
\begin{itemize}[leftmargin=*, itemsep=2pt, parsep=0pt, topsep=0pt, partopsep=0pt]
    \item End Users: who may be targeted by attackers. Since no private information is collected in this study, they face no direct harm. After publication, end users may benefit from the findings from this work via improved awareness.
    \item Organizations: whose brands are impersonated by attackers. We rely on both phishing data from official legal organizations and benign data from open websites. After publication, these organizations may experience reduced harm through better website design based on our findings.
    \item Researchers: who may adopt or extend our methods will gain insights into how to design the models.
    \item Companies: that develop and maintain large language models. We use open-source and commercially available LLMs, with no direct harm involved. They will benefit from the failure analysis of our work and obtain improved models after publication.
    \item Attackers: who may also gain some knowledge of LLMs; however, because these models are openly accessible, our contribution primarily strengthens defensive research rather than enabling attacks.
\end{itemize}

\PP{Impacts}We consider these principles: 

\begin{itemize}[leftmargin=*, itemsep=2pt, parsep=0pt, topsep=0pt, partopsep=0pt]
    \item Beneficence: Our final goal is to improve phishing detection and brand identification systems, reducing the risk of phishing.
    \item Respect for Persons: No human subjects were directly involved, and no personally identifiable information (PII) was collected or processed.
    \item Justice: We ensured balanced evaluation across multiple models (both commercial and open-source).
    \item Respect for Law and Public Interest: We follow the usage rules of the APWG eCX dataset, and the benign data are obtained from publicly available official websites. All models that we examined provide publicly accessible capabilities, whether open-source or commercial, under proper licensing or payment, and thus follow the law.  
\end{itemize}

\PP{Mitigations}Our work does not involve collecting private websites or exploiting users. A potential concern is the misuse of our datasets. To mitigate this risk, we ensure controlled access by requiring validated email verification through the dataset-sharing platform.

\PP{Decision}We determine that the benefits of this research outweigh the potential harms. By advancing understanding of how LLMs perform in phishing detection and by releasing curated, responsibly processed datasets, this work enables the community to design more interpretable, accurate, and practical defenses against phishing attacks. 

In conclusion, we attest that we consider the ethics of this research and believe that the research was done ethically, and our next-step plans are ethical.


\bibliographystyle{plain}
\bibliography{ref}

\appendix


\section{One-Shot Logo Input Example} ~\label{sec:oneshot_example}

\autoref{tab:prompt_example} presents a one-shot example using a ‘Chase’ logo to guide brand identification from logo inputs for LLMs. The blue highlights the structural components of the input.

\section{Configurations of Selected LLMs} ~\label{sec:appendix_llm_infor}

\autoref{tab:llm_info_sum} shows the configurations and settings of the selected LLMs, including input type, parameter size, execution location, organization, context length, base models, knowledge cutoff, and detailed model information.

\section{Categories of Detection Failure Reasons} ~\label{sec:addtional_conf_llms}

\autoref{tab:fail_explanation_category} provides a taxonomy of detection failure reasons with brief descriptions, constructed from \texttt{Gemini}’s summaries of explanations for misclassified cases.
\section{Word Distribution} ~\label{sec:total_word_distribution} 

\begin{table}[h]
\centering
\caption{\textbf{Categories of Detection Failure Reasons.}}
\label{tab:fail_explanation_category}
\resizebox{0.98\columnwidth}{!}{
    \begin{tabular}{p{0.325\columnwidth} p{0.65\columnwidth}}
    \toprule
    \multicolumn{1}{c}{\textbf{Category}} & \multicolumn{1}{c}{\textbf{Description}} \\ \midrule
    \multirow{2}{*}{Data Request} & Asking for sensitive user data (logins, financial, or personal information). \\\midrule
    \multirow{2}{*}{Insufficient} & Not enough reliable evidence or context to determine the site's purpose.\\\midrule
    \multirow{2}{*}{Mismatch} & Branding, layout, or code is inconsistent or mimics a legitimate service. \\\midrule
    \multirow{2}{*}{No Phishing Signals} & No technical, visual, or behavioral indicators of phishing attempts.\\\midrule
    \multirow{2}{*}{No Trust Signals} & Missing core security and transparency elements (HTTPS, policies, or contacts).\\ \midrule
    \multirow{2}{*}{Suspicious URLs} & Deceptive or abnormal domain/URL (typos, odd TLDs, shortened). \\\midrule
    \multirow{2}{*}{Trusted} & Trusted sites with consistent branding, policies, or verifiable ownership.\\
    
    \bottomrule
    \end{tabular}
}
\end{table}

\begin{table}[!t]
\centering
\caption{\textbf{One-Shot Logo Input Example for `Chase'.}}
\label{tab:prompt_example}
\renewcommand{\arraystretch}{1.2}
\begin{tabular}{p{0.95\linewidth}}
\centering
\includegraphics[width=0.2\linewidth]{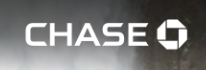} \\[4pt]
\rule{\linewidth}{0.4pt}
\raggedright
\colorbox{cyan!20}{\textbf{Task Information:}} 
Your task is to analyze the provided logos and determine the most likely brand. \\[4pt]

\colorbox{cyan!20}{\textbf{Answer Instruction:}} 
Replace the content in the \textless \textgreater with your answer. Strictly follow the Response Format:  
1. Brand: \textless Most Possible Brand Name\textgreater (If uncertain, state `Unknown').  
2. Evidence: \textless Provide Objective Evidence Within 100 Words\textgreater. \\[4pt]

\colorbox{cyan!20}{\textbf{One-Shot Example:}}  
Here is one example: The logo image is [input image], the answer is:  
1. Brand: Chase,  
2. Evidence: The text `CHASE' in uppercase is clearly visible, and the geometric shape consists of four trapezoidal segments arranged in a square-like formation with an open center, consistent with its brand.  

Here is the new logo image [new logo image]. How about this logo? \\[3pt]
\rule{\linewidth}{0.4pt}
\end{tabular}
\end{table}

\begin{table*}[!t]
\centering
\caption{\textbf{Overview of Our Selected LLMs and Configurations for Evaluation Experiments.}}
\label{tab:llm_info_sum}
\resizebox{0,98\linewidth}{!}{
\begin{tabular}{l r r r r r r r r}
\toprule
\textbf{} &  \multicolumn{1}{c}{\textbf{Input}} & \multicolumn{1}{c}{\textbf{Para.}}  & \multicolumn{1}{c}{\textbf{Run}} & \multicolumn{1}{c}{\textbf{Source}}   & \multicolumn{1}{c}{\textbf{Cont. Len.$^*$}} & \multicolumn{1}{c}{\textbf{Base}}  & \multicolumn{1}{c}{\textbf{Cutoff}} & \multicolumn{1}{c}{\textbf{Model/API}}\\ \midrule
{\textbf{GPT}~\cite{GPT:online}}       & Vision & NA    & API    &  Azure           & 1,047,576 & NA                  & 2024-05-31   & GPT-4.1                            \\ \midrule
{\textbf{Gemini}~\cite{Gemini:online}} & Vision & NA    & API    & Google Cloud     & 1,048,576 & NA                  & 2024-06      & gemini-2.0-flash-001               \\ \midrule
{\textbf{Qwen}~\cite{Qwen:online}}     & Vision & 8.29B & Local  & Hugging Face     & 8,192     & Qwen2.5             & NA           & Qwen2.5-VL-7B-Instruct             \\ \midrule
{\textbf{Janus}~\cite{Janus:online}}   & Vision & 7B    & Local  & Hugging Face     & 4,096     & DeepSeek-7b-base    & NA           & Janus-Pro-7B                       \\ \midrule
{\textbf{VL2}~\cite{VL2:online}}       & Vision & 1B    & Local  & Hugging Face     & 4,096     & DeepSeekMoE         & 2024-12      & deepseek-vl2-tiny                  \\ \midrule
{\textbf{Llama}~\cite{Llama:online}}   & Vision & 10.6B & Local  &  Ollama          & 131,072   & Llama 3.1           & 2023-12      & llama3.2-vision:11b-instruct-q8\_0 \\ \midrule
{\textbf{R1}~\cite{R1Distill:online}}  & Text   & 7B    & Local  &  vLLM            & 8,192     & Qwen2.5-Math-7B     & NA           & DeepSeek-R1-Distill-Qwen-7B        \\ \bottomrule
\multicolumn{6}{l}{$^*$Cont. Len. indicates the maximum token context window length.}
\end{tabular}
}
\end{table*}

\begin{table*}[!t]
\centering
\caption{\textbf{Categories of Explanation Words.}}
\label{tab:explanation_category}
\footnotesize
\resizebox{0.97\linewidth}{!}{
    \begin{tabular}{cl p{0.4\linewidth} p{0.3\linewidth}}
    \toprule
    \multicolumn{1}{c}{\textbf{Index}} & \multicolumn{1}{c}{\textbf{Category}} & \multicolumn{1}{c}{\textbf{Word Examples}} & \multicolumn{1}{c}{\textbf{Description}} \\ \midrule
    \multirow{2}{*}{1} & \multirow{2}{*}{Authentication \& Identity} & login, official, subdomain, account, user, email, password, security, identity, sign, identification, address. & Words related to user identity, login credentials, and verification mechanisms.\\\midrule
    \multirow{2}{*}{2} & \multirow{2}{*}{Layout \& Positioning} & medium, left, align, center, corner, site, structure, layout, footer, circle, shape, square, horizontal, circular, right. & Words related to spatial structure or positioning of elements on a webpage.\\\midrule
    \multirow{2}{*}{3} & \multirow{2}{*}{Color \& Style} & design, blue, color, white, font, scheme, stylize, background, lowercase, style, visual, red, green, bold, yellow. & Words related to visual appearance, including color schemes, typography, and overall style. \\\midrule
    \multirow{2}{*}{4} & \multirow{2}{*}{Interactive Elements} & support, link, form, button, interface, help, phone, option, request, mobile, navigation, search, telephone. & Words related to user-interactable components of a webpage. \\\midrule
    \multirow{2}{*}{5} & \multirow{2}{*}{Text \& Language} & text, content, title, name, message, term, language, phrase, keyword, copyright, english, mark, terminology. & Words related to textual content shown to the user, such as titles, phrases, or terms. \\\midrule
    \multirow{2}{*}{6} & \multirow{2}{*}{Media \& Graphics} & website, symbol, icon, web, tag, iconic, webpage, video, media, speech, imagery, trademark, photo. & Words related to visual media and imagery, such as icons.\\
    \bottomrule
    \end{tabular}
}
\end{table*}
\begin{figure}[t]
\centering
\includegraphics[width=.8\columnwidth]{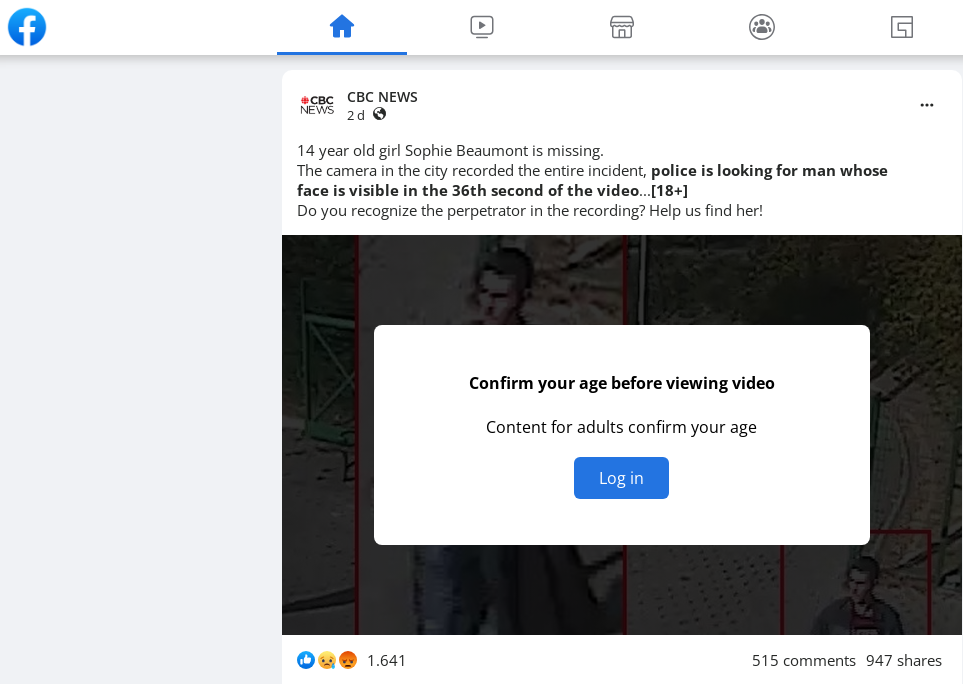}
\caption{\textbf{Screenshot Input Fail Case for GPT.}} 
~\label{fig:eg_scr_gpt_text}
\end{figure}

\begin{figure}[!t]
\centering
\includegraphics[width=0.35\columnwidth]{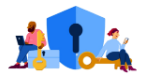}
\caption{\textbf{Logo Input Fail Case for GPT.}} 
~\label{fig:eg_logo_gpt_color}
\end{figure}

\begin{table*}[!t]
    \centering
    \caption{\textbf{Category Occurrence for 19,131 Samples on Various Input Prompting.} Values represent the count in this category and the corresponding percentage out of the total sample number. Bold values indicate a high percentage.}
    \label{tab:word_explanation}
    \resizebox{0.85\linewidth}{!}{
    \begin{tabular}{p{0.055\linewidth} p{0.06\linewidth} rrrrrr}
        \toprule
        \multirow{2}{*}{\textbf{Input}} & \multirow{2}{*}{\textbf{Model}} & \multicolumn{6}{c}{\textbf{Word Category}}\\
        \cmidrule(lr){3-8}
        & & \multicolumn{1}{c}{Auth.\&ID} & \multicolumn{1}{c}{Layout\&Pos.} & \multicolumn{1}{c}{Color\&Style} & \multicolumn{1}{c}{Inter. Element} & \multicolumn{1}{c}{Text\&Lang.} & \multicolumn{1}{c}{Media\&Graphics} \\\midrule
        \multirow{6}{*}{\textbf{Scr.}} 
        & GPT & \textbf{15,716 (82.15\%)} & 11,839 (61.88\%) & 11,007 (57.53\%) & 12,894 (67.40\%) & 14,842 (77.58\%) & 5,626 (29.41\%)\\
        & Gemini & 11,220 (58.65\%) & 6,422 (33.57\%) & 2,486 (12.99\%) & 9,701 (50.71\%) & \textbf{14,900 (77.88\%)} & 2,487 (13.00\%)\\
        & Qwen & 13,328 (69.67\%) & 13,359 (69.83\%) & 8,647 (45.20\%) & 11,157 (58.32\%) & \textbf{13,928 (72.80\%)} & 5,194 (27.15\%)\\
        & Llama  & \textbf{13,247 (69.24\%)} & 7,073 (36.97\%) & 7,939 (41.50\%) & 7,370 (38.52\%) & 10,186 (53.24\%) & 7,498 (39.19\%)\\
        & Janus & 13,853 (72.41\%) & 12,480 (65.23\%) & 12,693 (66.35\%) & 11,948 (62.45\%) & \textbf{14,987 (78.34\%)} & 7,960 (41.61\%)\\
        & VL2 & 8,305 (43.41\%) & 5,882 (30.75\%) & 2,310 (12.07\%) & 5,798 (30.31\%) & \textbf{8,678 (45.36\%)} & 3,213 (16.79\%)\\\midrule
        \multirow{6}{*}{\textbf{Logo}}  
        & GPT & 17,104 (89.40\%) & 13,540 (70.78\%) & \textbf{18,389 (96.12\%)} & 10,410 (54.41\%) & 13,339 (69.72\%) & 16,399 (85.72\%)\\
        & Gemini & 10,457 (54.66\%) & 8,940 (46.73\%) & \textbf{16,642 (86.99\%)} & 6,527 (34.12\%) & 11,626 (60.77\%) & 14,580 (76.21\%) \\
        & Qwen & 10,551 (55.15\%) & 10,270 (53.68\%) & \textbf{18,024 (94.21\%)} & 6,402 (33.46\%) & 14,713 (76.91\%) & 13,049 (68.21\%)\\
        & Llama & 13,968 (73.01\%) & 10,273 (53.70\%) & \textbf{15,966 (83.46\%)} & 7,426 (38.82\%) & 13,031 (68.11\%) & 11,810 (61.73\%)\\
        & Janus & 12,440 (65.03\%) & 12,829 (67.06\%) & \textbf{17,950 (93.83\%)} & 5,847 (30.56\%) & 14,634 (76.49\%) & 11,722 (61.27\%)\\
        & VL2 & 8,046 (42.06\%) & 10,089 (52.74\%) & \textbf{15,269 (79.81\%)} & 4,309 (22.52\%) & 13,853 (72.41\%) & 10,875 (56.84\%)\\\midrule
        \multirow{6}{*}{\textbf{URL}} 
        & GPT & \textbf{17,016 (88.94\%)} & 8,882 (46.43\%) & 5,618 (29.37\%) & 7,345 (38.39\%) & 9,812 (51.29\%) & 4,774 (24.95\%)\\
        & Gemini & \textbf{14,070 (73.55\%)} & 3,795 (19.84\%) & 2,210 (11.55\%) & 3,819 (19.96\%) & 8,000 (41.82\%) & 4,629 (24.20\%)\\
        & Qwen & 7,829 (40.92\%) & 7,895 (41.27\%) & 8,014 (41.89\%) & 9,802 (51.24\%) & \textbf{17,559 (91.78\%)} & 8,608 (45.00\%)\\
        & Llama & \textbf{16,934 (88.52\%)} & 5,496 (28.73\%) & 3,606 (18.85\%) & 6,943 (36.29\%) & 13,987 (73.11\%) & 8,009 (41.86\%)\\
        & Janus & 11,775 (61.55\%) & 3,601 (18.82\%) & 883 (4.62\%) & 3,870 (20.23\%) & \textbf{15,696 (82.04\%)} & 9,071 (47.42\%)\\
        & VL2 & 5,635 (29.45\%) & 4,415 (23.08\%) & 1,006 (5.26\%) & 3,391 (17.73\%) & \textbf{9,554 (49.94\%)} & 9,170 (47.93\%) \\
        & R1 & \textbf{7,581 (39.63\%)} & 5,291 (27.66\%) & 841 (4.40\%) & 2,730 (14.27\%) & 6,749 (35.28\%) & 2,764 (14.45\%) \\\midrule
        \multirow{6}{*}{\textbf{HTML}} 
        & GPT & 11,744 (61.39\%) & 9,372 (48.99\%) & 2,737 (14.31\%) & 11,959 (62.51\%) & \textbf{18,089 (94.55\%)} & 5,040 (26.34\%)\\
        & Gemini & 9,236 (48.28\%) & 8,105 (42.37\%) & 2,376 (12.42\%) & 10,000 (52.27\%) & \textbf{18,070 (94.45\%)} & 5,663 (29.60\%)\\
        & Qwen &13,313 (69.59\%) & 11,975 (62.59\%) & 4,524 (23.65\%) & 13,141 (68.69\%) & \textbf{17,904 (93.59\%)} & 10,810 (56.51\%)\\
        & Llama  & 10,129 (52.95\%) & 9,265 (48.43\%) & 4,624 (24.17\%) & 11,217 (58.63\%) & \textbf{17,119 (89.48\%)} & 9,988 (52.21\%)\\
        & Janus & 5,152 (26.93\%) & 3,401 (17.78\%) & 1,763 (9.22\%) & 4,351 (22.74\%) & \textbf{15,247 (79.70\%)} & 5,472 (28.60\%) \\
        & VL2 & 5,634 (29.45\%) & 5,345 (27.94\%) & 1,387 (7.25\%) & 6,499 (33.97\%) & \textbf{10,982 (57.40\%)} & 9,035 (47.23\%) \\
        & R1 & 9,697 (50.69\%) & 7,678 (40.13\%) & 1,457 (7.62\%) & 10,735 (56.11\%) & \textbf{16,217 (84.77\%)} & 6,346 (33.17\%)\\\midrule
        \multirow{6}{*}{\makecell[l]{\textbf{Scr.}\\\textbf{URL}}} 
        & GPT & \textbf{18,447 (96.42\%)} & 13,641 (71.30\%) & 14,821 (77.47\%) & 15,000 (78.41\%) & 16,832 (87.98\%) & 6,644 (34.73\%)\\
        & Gemini & 13,315 (69.60\%) & 5,169 (27.02\%) & 2,104 (11.00\%) & 11,021 (57.61\%) & \textbf{14,866 (77.71\%)} & 3,057 (15.98\%) \\
        & Qwen & 15,502 (81.03\%) & 13,130 (68.63\%) & 6,451 (33.72\%) & 11,476 (59.99\%) & \textbf{16,391 (85.68\%)} & 9,965 (52.09\%) \\
        & Llama & 12,256 (64.06\%) & 8,182 (42.77\%) & 6,512 (34.04\%) & 9,954 (52.03\%) & \textbf{12,257 (64.07\%)} & 8,311 (43.44\%) \\
        & Janus & 5,274 (27.57\%) & 2,240 (11.71\%) & 2,010 (10.51\%) & 3,255 (17.01\%) & \textbf{8,222 (42.98\%)} & 2,969 (15.52\%) \\
        & VL2 & 4,477 (23.40\%) & 2,252 (11.77\%) & 1,478 (7.73\%) & 3,002 (15.69\%) & \textbf{6,094 (31.85\%)} & 3,033 (15.85\%) \\ \midrule
        \multirow{6}{*}{\makecell[l]{\textbf{Logo} \\\textbf{URL}}} 
        & GPT & \textbf{18,279 (95.55\%)} & 14,755 (77.13\%) & 17,972 (93.94\%) & 11,886 (62.13\%) & 15,551 (81.29\%) & 14,206 (74.26\%)\\
        & Gemini & 10,213 (53.38\%) & 6,253 (32.69\%) & 10,569 (55.25\%) & 6,219 (32.51\%) & \textbf{11,104 (58.04\%)} & 10,463 (54.69\%)\\
        & Qwen & 14,609 (76.36\%) & 13,756 (71.90\%) & 15,620 (81.65\%) & 13,693 (71.57\%) & \textbf{16,401 (85.73\%)} & 15,420 (80.60\%)\\
        & Llama & \textbf{14,921 (77.99\%)} & 11,085 (57.94\%) & 12,799 (66.90\%) & 11,855 (61.97\%) & 14,408 (75.31\%) & 12,685 (66.31\%) \\
        & Janus & 10,140 (53.00\%) & 12,680 (66.28\%) & \textbf{14,710 (76.89\%)} & 5,893 (30.80\%) & 13,525 (70.70\%) & 11,485 (60.03\%)\\
        & VL2 & 8,683 (45.39\%) & 9,999 (52.27\%) & 9,953 (52.03\%) & 6,079 (31.78\%) & \textbf{12,622 (65.98\%)} & 11,763 (61.49\%)\\ \midrule
        \multirow{5}{*}{\makecell[l]{\textbf{Scr.}\\\textbf{URL}\\\textbf{HTML}}} 
        & GPT & 18,542 (96.92\%) & 14,384 (75.19\%) & 13,363 (69.85\%) & 16,115 (84.24\%) & \textbf{18,977 (99.20\%)} & 7,231 (37.80\%)\\
        & Gemini & 12,795 (66.88\%) & 6,902 (36.08\%) & 2,231 (11.66\%) & 11,716 (61.24\%) & \textbf{18,512 (96.76\%)} & 4,314 (22.55\%)\\
        & Qwen & 15,835 (82.77\%) & 15,100 (78.93\%) & 7,462 (39.00\%) & 14,883 (77.80\%) & \textbf{18,949 (99.05\%)} & 12,073 (63.11\%)\\
        & Llama & 13,238 (69.20\%) & 10,136 (52.98\%) & 6,273 (32.79\%) & 13,982 (73.09\%) & \textbf{18,459 (96.49\%)} & 15,398 (80.49\%) \\
        & Janus & 8,334 (43.56\%) & 3,959 (20.69\%) & 3,291 (17.20\%) & 6,905 (36.09\%) & \textbf{17,138 (89.58\%)} & 13,514 (70.64\%)\\
        & VL2 & 6,069 (31.72\%) & 3,369 (17.61\%) & 1,562 (8.16\%) & 4,666 (24.39\%) & \textbf{7,628 (39.87\%)} & 4,696 (24.55\%)\\ \midrule
        \multirow{5}{*}{\makecell[l]{\textbf{Scr.}\\\textbf{One-shot}}}
        & GPT & \textbf{15,102 (78.94\%)} & 10,430 (54.52\%) & 11,707 (61.19\%) & 13,521 (70.68\%) & 13,434 (70.22\%) & 3,519 (18.39\%)\\
        & Gemini & \textbf{13,341 (69.73\%)} & 6,465 (33.79\%) & 4,115 (21.51\%) & 12,936 (67.62\%) & 12,321 (64.40\%) & 2,627 (13.73\%)\\
        & Qwen & 16,490 (86.20\%) & 15,588 (81.48\%) & 8,240 (43.07\%) & \textbf{16,883 (88.25\%)} & 15,195 (79.43\%) & 6,714 (35.09\%)\\
        & Janus & 12,956 (67.72\%) & 8,499 (44.43\%) & 13,445 (70.28\%) & \textbf{13,505 (70.59\%)} & 8,879 (46.41\%) & 4,875 (25.48\%)\\
        & VL2 & \textbf{16,716 (87.38\%)} & 4,644 (24.27\%) & 2,915 (15.24\%) & 13,514 (70.64\%) & 3,502 (18.31\%) & 979 (5.12\%)\\ \midrule
        \multirow{5}{*}{\makecell[l]{\textbf{Logo}\\\textbf{One-shot}}} 
        & GPT & 14,305 (74.77\%) & 11,959 (62.51\%) & \textbf{18,500 (96.70\%)} & 7,639 (39.93\%) & 12,988 (67.89\%) & 15,802 (82.60\%)\\
        & Gemini & 7,995 (41.79\%) & 9,830 (51.38\%) & \textbf{17,439 (91.16\%)} & 4,819 (25.19\%) & 11,618 (60.73\%) & 13,037 (68.15\%)\\
        & Qwen & 11,683 (61.07\%) & 14,627 (76.46\%) & \textbf{18,502 (96.71\%)} & 8,720 (45.58\%) & 15,160 (79.24\%) & 14,192 (74.18\%)\\
        & Janus & 2,858 (14.94\%) & 14,962 (78.21\%) & \textbf{18,292 (95.61\%)} & 2,942 (15.38\%) & 18,089 (94.55\%) & 7,543 (39.43\%)\\
        & VL2 & 4,020 (21.01\%) & 15,068 (78.76\%) & \textbf{17,734 (92.70\%)} & 2,326 (12.16\%) & 8,492 (44.39\%) & 13,902 (72.67\%)\\ \midrule
        \multirow{7}{*}{\makecell[l]{\textbf{URL}\\\textbf{One-shot}}} 
        & GPT & \textbf{17,234 (90.08\%)} & 8,203 (42.88\%) & 6,442 (33.67\%) & 7,498 (39.19\%) & 13,295 (69.49\%) & 4,882 (25.52\%)\\
        & Gemini & \textbf{14,581 (76.22\%)} & 5,995 (31.34\%) & 3,226 (16.86\%) & 5,197 (27.17\%) & 10,342 (54.06\%) & 6,141 (32.10\%)\\
        & Qwen & 8,644 (45.18\%) & 8,810 (46.05\%) & 11,264 (58.88\%) & 5,778 (30.20\%) & \textbf{15,410 (80.55\%)} & 7,326 (38.29\%)\\
        & Llama & \textbf{16,443 (85.95\%)} & 7,837 (40.96\%) & 4,055 (21.20\%) & 7,088 (37.05\%) & 13,902 (72.67\%) & 7,322 (38.27\%)\\
        & Janus & 16,932 (88.51\%) & 3,368 (17.60\%) & 10,549 (55.14\%) & 3,133 (16.38\%) & \textbf{16,966 (88.68\%)} & 4,737 (24.76\%) \\
        & VL2 & 6,772 (35.40\%) & 4,923 (25.73\%) & 973 (5.09\%) & 2,589 (13.53\%) & \textbf{9,406 (49.17\%)} & 8,809 (46.05\%)\\ 
        & R1 & \textbf{6,896 (36.05\%)} & 4,595 (24.02\%) & 924 (4.83\%) & 2,148 (11.23\%) & 4,961 (25.93\%) & 2,099 (10.97\%) \\ \bottomrule
    \end{tabular}
    }
\end{table*}

In this section, we discuss the word distribution of all LLM responses with temperature at 0.0/0.1. Specifically, we summarize the categories and their descriptions in \autoref{tab:explanation_category}. We discuss the distribution of the single component in \aref{subsec:depth_single_total}, combinations in \aref{subsec:depth_combination_total}, one-shot in \aref{subsec:depth_oneshot_total}, and provide two examples in \aref{subsec:example}.

\subsection{Analysis of Single Component} \label{subsec:depth_single_total}

Among the screenshot inputs, all models demonstrate strong activation across multiple categories, especially in \textit{Text \& Language} and \textit{Authentication \& Identity}. \texttt{GPT} and \texttt{Llama} detect 82.15\% and 69.24\% of samples in the \textit{Authentication \& Identity}, while \texttt{Gemini}, \texttt{Qwen}, \texttt{Janus}, and \texttt{VL2} detect 77.88\%, 72.80\%, 78.34\%, and 45.36\% of samples in the \textit{Text \& Language} category, respectively. These results highlight the richness and diversity of semantic information present in screenshots.
All models show a strong preference for \textit{Color \& Style} with logo inputs, with high percentages from \texttt{GPT} (96.12\%), \texttt{Gemini} (86.99\%), \texttt{Qwen} (94.21\%), \texttt{Llama} (83.46\%), \texttt{Janus} (93.83\%), and \texttt{VL2} (79.81\%). Compared to screenshot inputs, models also demonstrate notable attention to \textit{Text \& Language}, but the proportion of \textit{Color \& Style} is higher, indicating the visual nature of logos.
URL inputs primarily activate \textit{Authentication \& Identity} and \textit{Text \& Language} due to their textual and identity-oriented content. 
HTML inputs consistently yield the preference for \textit{Text \& Language}, with \texttt{GPT} (94.55\%), \texttt{Gemini} (94.45\%), \texttt{Qwen} (93.59\%), \texttt{Llama} (89.48\%), \texttt{Janus} (79.70\%), \texttt{VL2} (57.40\%), \texttt{R1} (84.77\%), respectively, but lower \textit{Media \& Graphics} and \textit{Color \& Style} detection compared to screenshot and logo inputs.

\observ{The type of input prompts significantly affects the kind of information LLMs extract. Screenshot inputs, visual and textual rich, enable LLMs to generate well-rounded outputs but show a preference for \textit{Authentication \& Identity} and \textit{Text \& Language}. Logo inputs emphasize \textit{Color \& Style}, URL inputs are limited to \textit{Authentication \& Identity} and \textit{Text \& Language}, while HTML inputs also prefer \textit{Text \& Language} due to their predominantly textual nature}

\subsection{Analysis of Combination Results} \label{subsec:depth_combination_total}
Screenshot-URL inputs show the words cover multiple categories, especially in \textit{Text \& Language} and \textit{Authentication \& Identity}. Compared to screenshot inputs, \texttt{GPT} (10.40\%{\small\faLongArrowUp}), \texttt{Qwen} (12.88\%{\small\faLongArrowUp}), and \texttt{Llama} (10.82\%{\small\faLongArrowUp}) show improved coverage for \textit{Text \& Language}, while \texttt{Gemini} (0.17\%{\small\faLongArrowDown}), \texttt{Janus} (35.36\%{\small\faLongArrowDown}), and \texttt{VL2} (13.51\%{\small\faLongArrowDown}) show decreases. For \textit{Text \& Language}, \texttt{GPT} (14.27\%{\small\faLongArrowUp}), \texttt{Gemini} (10.95\%{\small\faLongArrowUp}), and \texttt{Qwen} (11.36\%{\small\faLongArrowUp}) demonstrate increases, whereas \texttt{Llama} (5.18\%{\small\faLongArrowDown}), \texttt{Janus} (44.84\%{\small\faLongArrowDown}), and \texttt{VL2} (20.01\%{\small\faLongArrowDown}) shows declines.
The addition of URLs pushes the coverage change.

\subsection{Analysis of One-shot Results} \label{subsec:depth_oneshot_total}
Among the screenshot-one-shot inputs, LLMs primarily focus on \textit{Authentication \& Identity} and \textit{Interactive Element}, in contrast to the screenshot inputs, which are dominated by \textit{Authentication \& Identity} and \textit{Text \& Language}. In the \textit{Authentication \& Identity} category, \texttt{GPT} (3.21\%{\small\faLongArrowDown}) and \texttt{Janus} (4.69\%{\small\faLongArrowDown}) show decreased coverage, whereas \texttt{Gemini} (11.08\%{\small\faLongArrowUp}), \texttt{Qwen} (16.53\%{\small\faLongArrowUp}), and \texttt{VL2} (43.97\%{\small\faLongArrowUp}) indicate increases. These trends suggest that one-shot guidance influences screenshot-based brand identification for LLMs, though the effect varies across LLMs.

\subsection{Examples of Single Component Failure} \label{subsec:example}
We present two examples of failures for screenshot and logo inputs in~\autoref{fig:eg_scr_gpt_text} and~\autoref{fig:eg_logo_gpt_color}, respectively. The original target brands are both \textit{Facebook}, but the screenshot is detected as \textit{CBC News} while the logo is recognized as \textit{Microsoft}.

\end{document}